\documentclass[11pt]{article}
\usepackage{amssymb}
\usepackage{amsmath}
\usepackage{latexsym}
\catcode`@=11
\renewcommand{\section}{\@startsection{section}{1}{0pt}{\medskipamount}
{\medskipamount}{\large\bf}}
\numberwithin{equation}{section}
\catcode`@=12

\def\th{\theta}

\def\g{\gamma}
\def\de{\delta}
\def\e{\epsilon}

\def\o{\omega}

\def\vp{\varphi}
\def\vt{\vartheta}
\def\p{\phi}
\def\s{\sigma}

\def\sfrac#1#2{{\textstyle\frac{#1}{#2}}}
\def\m{\mu}
\def\n{\nu}
\def\pa{\partial}

\renewcommand{\e}{\,\mathrm{e}\,}
\newcommand{\im}{\,\mathrm{i}\,}
\newcommand{\diff}{\mathrm{d}}
\newcommand{\rc}{{\mathbb{R}^{2n}}}

\newcommand{\rts}{{\mathbb{R}^{2n}_\theta{\times}S^2}}
\newcommand{\rt}{{\mathbb{R}^{2n}_\theta}}

\newcommand{\rns}{{\mathbb{R}^{2n}{\times}S^2}}
\newcommand{\R}{{\mathbb{R}}}
\newcommand{\C}{{\mathbb{C}}}
\newcommand{\Z}{{\mathbb{Z}}}
\newcommand{\Q}{{\mathbb{Q}}}
\newcommand{\Id}{\mathbf{1}_k}
\newcommand{\Idd}{\mathbf{1}}
\newcommand{\Hcal}{{\cal H}}

\newcommand{\Bcal}{{\cal B}}
\newcommand{\Kcal}{{\cal K}}
\newcommand{\Ncal}{{\cal N}}
\newcommand{\fh}{\hat{f}}

\newcommand{\xh}{\hat{x}}
\newcommand{\zh}{\hat{z}}
\newcommand{\zbh}{\hat{\bar{z}}}
\newcommand{\yb}{\bar{y}}
\newcommand{\zb}{\bar{z}}
\newcommand{\ab}{\bar{a}}
\newcommand{\bb}{\bar{b}}
\newcommand{\cb}{\bar{c}}
\newcommand{\ca}{{\cal{A}}}
\newcommand{\cf}{{\cal{F}}}
\newcommand{\cliff}{{{\rm C}\ell}}
\newcommand{\K}{{\sf K}}
\newcommand{\Ka}{{\sf K}^{\rm a}}
\newcommand{\Kt}{{\sf K}^{\rm t}}
\newcommand{\ch}{{\mathrm{ch}}}
\newcommand{\Lcal}{{{\cal L}^{~}_{\rm D}}}
\newcommand{\Tr}[1]{\:{\rm Tr}\,#1}
\newcommand{\mbf}[1]{{\boldsymbol {#1} }}

\def\Dirac{{D\!\!\!\!/\,}} 
\def\mDelta{{\mbf\Delta}}
\def\msphere{{\mbf\Sigma}}
\def\mball{{\mbf B}}
\def\>{\rangle}
\def\<{\langle}
\def\+{\dagger}
\def\={\ =\ }

\begin{document}
\begin{titlepage}
\setcounter{page}{0}
\begin{flushright}
hep-th/0310267\\
ITP--UH--08/03\\
HWM--03--24\\
EMPG--03--20
\end{flushright}

\vskip 1.8cm

\begin{center}

{\Large\bf Noncommutative Instantons in Higher Dimensions,}

{\Large\bf Vortices and Topological K-Cycles}

\vspace{15mm}

{\large Olaf Lechtenfeld${}^1$}, {\large Alexander D. Popov${}^{1,2}$}
\ \ and \ \ {\large Richard J. Szabo${}^3$}
\\[5mm]
\noindent ${}^1${\em Institut f\"ur Theoretische Physik,
Universit\"at Hannover \\
Appelstra\ss{}e 2, 30167 Hannover, Germany }
\\[5mm]
\noindent ${}^2${\em Bogoliubov Laboratory of Theoretical Physics, JINR\\
141980 Dubna, Moscow Region, Russia}
\\[5mm]
\noindent ${}^3${\em Department of Mathematics, Heriot-Watt University\\
Scott Russell Building, Riccarton, Edinburgh EH14 4AS, U.K.}
\\[5mm]
{Email: {\tt lechtenf, popov @itp.uni-hannover.de , R.J.Szabo@ma.hw.ac.uk}}

\vspace{15mm}

\begin{abstract}
\noindent

We construct explicit BPS and non-BPS solutions of the ${\rm U}(2k)$
Yang-Mills equations on the noncommutative space
$\mathbb{R}^{2n}_\theta{\times}S^2$ with finite energy and topological
charge. By twisting with a Dirac multi-monopole bundle over $S^2$, we
reduce the Donaldson-Uhlenbeck-Yau equations on
$\mathbb{R}^{2n}_\theta{\times}S^2$ to vortex-type equations for a
pair of ${\rm U}(k)$ gauge fields and a bi-fundamental scalar field on
$\mathbb{R}^{2n}_\theta$. In the ${\rm SO}(3)$-invariant case the
vortices on $\mathbb{R}^{2n}_\theta$ determine multi-instantons on
$\mathbb{R}^{2n}_\theta{\times}S^2$. We show that these solutions give
natural physical realizations of Bott periodicity and vector bundle
modification in topological K-homology, and can be interpreted as a
blowing-up of D0-branes on $\mathbb{R}^{2n}_\theta$ into spherical
D2-branes on $\mathbb{R}^{2n}_\theta{\times}S^2$. In the generic case
with broken rotational symmetry, we argue that the D0-brane charges on
$\mathbb{R}^{2n}_\theta{\times}S^2$ provide a physical interpretation
of the Adams operations in K-theory.

\end{abstract}

\end{center}
\end{titlepage}

\section{ Introduction }

It is a major enterprise of mathematical physics to try to extend the success
story of gauge field theory by adding supersymmetry or extra dimensions,
by embedding it into string theory, by deforming it noncommutatively,
or by combinations of these ideas. Specifically, Gr\"onewold-Moyal
type noncommutative deformations have been the subject
of intense research in recent years~\cite{Seiberg,Harvey}.
Besides settling the issue of perturbative quantization, it is important
to map out the classical configuration space of noncommutative gauge
theories and to characterize its role in string theory.
All celebrated BPS~configurations, such as instantons~\cite{Belavin},
monopoles~\cite{BPS} and vortices~\cite{Nielsen, Taubes}, have been generalized
to the noncommutative case, originally in~\cite{NS},~\cite{GN}
and~\cite{vortex,Bak},\footnote{
For related works on flux tube solutions see \cite{Poly,GN2}.}
respectively (see~\cite{Haman} for a recent review
and references). The description of D-branes as solitons in open
string field theory simplifies dramatically in the context of a
noncommutative tachyon field theory, with the D-branes appearing as
noncommutative solitons~\cite{DMR1}. The relation between D-branes
and noncommutative tachyons makes manifest~\cite{Matsuo1,HM1} the
relationship between D-branes and K-theory~\cite{MM1,OS1}.

In the superstring theory context, one encounters gauge theories in
spacetime dimensionalities up to ten. Already 20 years ago, BPS-type
equations in more than four dimensions were proposed~\cite{Corrigan2,
  Ward} and their solutions investigated e.g. in~\cite{Ward, Fairlie}.
More recently, noncommutative instantons in higher dimensions and their 
brane interpretations have been considered in~\cite{Witten}--\cite{Nekrasov}.
For nonabelian gauge theory on a K\"ahler manifold the most natural
BPS~condition lies in the Donaldson-Uhlenbeck-Yau
equations~\cite{Donaldson,Uhlenbeck}, which generalize the
four-dimensional self-duality equations. 

In this paper we investigate the Donaldson-Uhlenbeck-Yau equations on
the noncommutative spaces $\R^{2n}_{\th}{\times}S^2$ for the gauge
group U($2k$). By employing a reduction via a Dirac monopole bundle
over the $S^2$ we obtain generalized coupled vortex equations on
$\R^{2n}_{\th}$ for a pair of U($k$) gauge fields and a complex 
matrix-valued scalar field. Invoking partial isometries and the ABS 
construction, BPS and also non-BPS solutions are found, which are 
labelled by three integers and carry a number of moduli. We calculate 
the topological charge and energy of these ``multi-instanton'' solutions 
and mention some extremal cases. We then address the problem of 
assigning K-theory classes to the explicit noncommutative 
instanton solutions that are found. We will find that in the simplest 
instance of a one-monopole configuration on the $S^2$ 
our solutions provide physical realizations of the vector bundle 
modification relation in topological K-homology, and thereby yield 
explicit K-cycle representatives. For multi-monopole configurations 
on the $S^2$, we argue that the solutions instead realize symmetry
operations in K-theory. Using these correspondences we argue that in
the former case our instanton solutions describe D2-branes on $\rts$
which are equivalent to D0-branes on $\rt$ described by vortices. In
the latter case, the brane interpretations are less transparent, and
we argue that the vortex solutions represent D0-branes on $\rts$ which
contain residual moduli from their locations inside the $S^2$.

The organisation of this paper is as follows. In section~2 we recall
various aspects of the noncommutative space $\rts$, and in section~3
we write down the Donaldson-Uhlenbeck-Yau equations on it. In
section~4 we describe our particular ansatz that we use to solve these
equations, and show in section~5 how together they reduce to vortex
equations on $\rt$. Sections~6 and~7 then deal with explicit BPS and
non-BPS solutions, respectively, to the Yang-Mills equations on
$\rts$. In section~8 the topological charge of these configurations is
computed, and the Yang-Mills action is evaluated on them in
section~9. In section~10 solutions in the zero monopole and zero
tachyon sectors are explicitly constructed. In section~11 we show that
our solutions naturally define K-cycles, and this correspondence is
exploited in section~12 to assign D0-brane charges to them. Finally,
in section~13 we summarize our findings and make some further remarks
concerning the brane interpretations of the multi-instanton solutions
on $\rts$.

\section{The noncommutative space $\mbf{\rts}$ }

\noindent
{\bf Geometry of $\mbf{\rns}$.\ }
In order to set the stage for the Donaldson-Uhlenbeck-Yau
equations and their instanton solutions, we begin with
the (commutative) manifold $\rns$ carrying the Riemannian metric
\begin{equation}\label{metric}
\diff s^2\= \de_{\mu\nu}~\diff x^\mu~\diff x^\nu
+ R^2\left(\diff \vt^2 + \sin^2\vt\ \diff\vp^2\right)
\=  g_{ij}~\diff x^i~\diff x^j\ ,
\end{equation}
where $\m,\n=1,\ldots,2n$ but $i,j=1,\ldots,2n{+}2$, and
$x=(x^\mu )$ are coordinates on $\rc$ while
$x^{2n+1}{=}\vt$ and $x^{2n+2}{=}\vp$ parametrize the
standard two-sphere $S^2$ of constant radius~$R$,
i.e.~$0\le\vp\le 2\pi$ and $0\le\vt\le \pi$.
We use the Einstein summation convention for repeated indices.
The volume two-form on $S^2$ reads
\begin{equation}\label{volume}
\sqrt{\det(g_{ij})}\ \diff\vt\wedge\diff\vp
\ =:\ \o_{\vt\vp}\ \diff\vt\wedge\diff\vp \= \o
\qquad\Longrightarrow\qquad
\o_{\vt\vp} = -\o_{\vp\vt}= R^2\sin\vt\ .
\end{equation}
The manifold $\rns$ is K\"ahler,
with local complex coordinates $z^1,\ldots,z^n,y$ where
\begin{equation}\label{zz}
z^a\=x^{2a-1}-\im\,x^{2a} \qquad\textrm{and}\qquad
\zb^{\ab}\=x^{2a-1}+\im\,x^{2a} \qquad\textrm{with}\quad
a=1,\ldots,n
\end{equation}
and
\begin{equation}\label{zn1}
y\=\frac{R\,\sin\vt}{(1+\cos\vt )}\,\exp{(-\im\vp)}\quad ,\qquad
\yb\=\frac{R\,\sin\vt}{(1+\cos\vt) }\,\exp{(\im\vp)}\ ,
\end{equation}
so that $1{+}\cos\vt=\frac{2R^2}{R^2{+}y\yb}$.
In these coordinates, the metric takes the form
\begin{equation}
\diff s^2 \= \de_{a\bb}\ \diff z^a~\diff\zb^{\bb}
+ \sfrac{4\,R^4}{(R^2+y\yb)^2}~\diff y~\diff\yb
\end{equation}
with $\de_{a\ab}{=}\de^{a\ab}{=}1$ (and all other entries vanishing),
and the K\"ahler two-form reads
\begin{equation}\label{kahler}
\Omega\=-\sfrac{{\im}}{2}\,\bigl\{
\delta_{a\bb}\ \diff z^a\wedge\diff \zb^{\bb} +
\sfrac{4\,R^4}{(R^2+y\yb)^2}\ \diff y\wedge\diff \yb\bigr\}
\=-\sfrac{{\im}}{2}\,\de_{a\bb}\ \diff z^a\wedge\diff \zb^{\bb} +
\o_{\vt\vp}\ \diff \vt\wedge\diff \vp \ .
\end{equation}
For later use, we also note here the derivatives
\begin{equation}\label{pazz}
\pa_{z^a} \= \sfrac{1}{2}\,(\pa_{2a-1}+ \im \pa_{2a}) \qquad\textrm{and}\qquad
\pa_{\zb^{\ab}} \= \sfrac{1}{2}\,(\pa_ {2a-1} - \im \pa_{2a})\ ,
\end{equation}
where $\pa_{\m}:={\pa}/\pa {x^\m}$ for $\mu{=}1,\ldots,2n$.

\bigskip

\noindent
{\bf Noncommutative deformation.\ }
Let us now pass to a noncommutative deformation of the flat part of the
manifold under consideration, i.e. $\rns\to\rts$.
Note that the $S^2$ factor remains commutative in this paper.
As is well known, classical field theory on the noncommutative space
$\R^{2n}_\th$ may be realized in either a star-product formulation
or in an operator formalism.
While the first approach alters the product of functions on~$\R^{2n}$,
the second one turns these functions~$f$ into linear operators~$\fh$
acting on the $n$-harmonic oscillator Fock space~$\Hcal$.
The noncommutative space~$\R^{2n}_\th$ may then be defined by declaring
its coordinate functions $\xh^1,\ldots,\xh^{2n}$ to obey
the Heisenberg algebra relations
\begin{equation}
[ \xh^\mu\,,\,\xh^\nu ] \= \im\,\th^{\mu\nu}
\end{equation}
with a constant antisymmetric tensor~$\th^{\mu\nu}$.
The coordinates can be chosen in such a way that the matrix
$\theta=(\th^{\m\n})$ is block-diagonal with non-vanishing components
\begin{equation}\label{tha}
\th^{{2a-1}\ {2a}} \= -\th^{{2a}\ {2a-1}} \ =:\ \th^a \ .
\end{equation}
We will assume that all $\th^a\ge0$, as
the general case does not hide additional complications.
For the noncommutative version of the complex coordinates (\ref{zz}) we have
\begin{equation}\label{zzb}
\big[\zh^a\,,\,\zbh^{\bb}\big] \= -2\,\de^{a\bb}\,\th^a \
=:\ \th^{a\bb} \= -\th^{\bb a}\ \le 0
\quad,\qquad \mbox{and all other commutators vanish}\ .
\end{equation}
The Fock space~$\Hcal$ is spanned by the basis states
\begin{equation}
|k_1,\ldots,k_n\>\=\prod_{a=1}^{n}(2\th^a\,k_a!)^{-1/2}\,(\zh^{a})^{k_a}
|0,\ldots,0\>
\qquad \textrm{for} \quad k_a=0,1,2,\ldots \ ,
\end{equation}
which are connected by the action of creation and annihilation operators
subject to the commutation relations
\begin{equation}
\Bigl[\,\frac{\zbh^{\bb}}{\sqrt{2\th^b}}\ ,\ \frac{\zh^a}{\sqrt{2\th^a}}\,
\Bigr] \= \de^{a\bb} \ .
\end{equation}

We recall that, in the Weyl operator realization $f{\mapsto}\fh$,
derivatives of a function~$f$ on $\R^{2n}$ get mapped according to
\begin{equation}\label{pazzf}
\widehat{\pa_{z^a} f}\=\th_{a\bb}\,\big[\zbh^{\bb} \,,\, \fh\,\big]
\ =:\ \pa_{\zh^a} \fh
\qquad\textrm{and}\qquad
\widehat{\pa_{\zb^{\ab}} f}\=\th_{\ab b}\,\big[\zh^b \,,\, \fh\,\big]
\ =:\ \pa_{\zbh^{\ab}} \fh\ ,
\end{equation}
where $\th_{a\bb}$ is defined via $\th_{b\bar{c}}\,\th^{\bar{c}a}=\de^a_b$
so that $\th_{a\bb}=-\th_{\bb a}=\frac{\de_{a\bb}}{2\th^a}$.
Finally, there is the relation
\begin{equation}
\int_{\R^{2n}} \!\diff^{2n} x~f(x)\=
\Bigl( \prod_{a=1}^n 2\pi\th^a \Bigr)~\textrm{Tr}^{~}_\Hcal\,\fh\ .
\end{equation}
Taking the product of $\R^{2n}_\th$ with the commutative sphere $S^2$
means extending the noncommutativity matrix $\th$ by vanishing entries
in the two new directions. A more detailed description of
noncommutative field theories can be found in~\cite{Harvey}.

\section{The Donaldson-Uhlenbeck-Yau equations }

The generalization of the four-dimensional self-duality equations to
higher dimensions is not unique. A particularly natural extension is given
by the Donaldson-Uhlenbeck-Yau (DUY) equations~\cite{Donaldson,Uhlenbeck}
which can be formulated on any K\"ahler manifold. Their importance derives
from the BPS property, i.e. they yield stable solutions of the Yang-Mills
equations.
We shall present the DUY equations first in generality, then on $\rns$,
and finally on $\rts$.

Let $M_{2q}$ be a complex $q{=}n{+}1$ dimensional K\"ahler manifold
with some local real coordinates $(x^i)$ and a tangent space basis
$\pa_{i} :=\pa /\pa x^i$ for $i,j =1,\ldots,2q$, so that the metric and
K\"ahler two-form read $\diff s^2=g_{ij}~\diff x^i~\diff x^j$ and
$\Omega=\Omega_{ij}~\diff x^i\wedge\diff x^j$, respectively.
Consider a rank~$2k$ complex vector bundle over~$M_{2q}$ with
a chosen gauge potential ${\ca}={\ca}_{i}~\diff x^i$.
The curvature two-form $\cf =\diff {\ca}+{\ca}\wedge {\ca}$ has components
${\cf}_{ij}=\pa_{i}{\ca}_{j}-\pa_{j}{\ca}_{i}+[{\ca}_{i},{\ca}_{j}]$
and the K\"ahler decomposition ${\cf}={\cf}^{2,0}+{\cf}^{1,1}+{\cf}^{0,2}$.
Both ${\ca}_{i}$ and ${\cf}_{ij}$ take values in the Lie algebra $u(2k)$.
The DUY equations~\cite{Donaldson,Uhlenbeck} on $M_{2q}$ are
\begin{equation}\label{DUY}
*\Omega\wedge {\cf}\ =\ 0 \qquad\textrm{and}\qquad
{\cf}^{0,2}\=0\={\cf}^{2,0}\ ,
\end{equation}
where
$*$ is the Hodge duality operator.
In our local coordinates $(x^i)$ we have
$q!\,(*\Omega\wedge{\cf})=(\Omega,{\cf})\,\Omega^q=\Omega^{ij}{\cf}_{ij}\,
\Omega^q$
where $\Omega^{ij}$ are defined via $\Omega^{ij}\,\Omega_{jl}=\de^i_l$.
For $q{=}2$ the DUY equations (\ref{DUY}) coincide with the anti-self-dual
Yang-Mills (ASDYM) equations
\begin{equation}\label{sdym1}
* {\cf} \ =\ - {\cf}
\end{equation}
introduced in~\cite{Belavin}.

Specializing now $M_{2q}$ to be $\rns$, the DUY equations (\ref{DUY})
in the complex local coordinates $(z^a,y)$ take the form\footnote{Note that
these equations are not integrable even for $n=1$. Therefore, neither
the dressing nor splitting approaches developed in~\cite{LP1} for equations
on noncommutative spaces can be applied here. The modified ADHM
construction~\cite{NS} also does not work in this case.}
\begin{align}\label{DU1}
\de^{a\bb}\,{\cf}_{z^a\zb^{\bb}}+
\sfrac{(R^2+y\yb)^2}{4\,R^4}\,{\cf}_{y\yb}&\=0\ ,\\
\label{DU2}
{\cf}_{\zb^{\ab}\zb^{\bb}}&\=0\={\cf}_{z^a z^b}\ ,\\
\label{DU3}
{\cf}_{\zb^{\ab}\yb}&\=0\={\cf}_{z^a y}\ ,
\end{align}
where $a,b=1,\ldots,n$.
Using the formulae (\ref{zn1}), we obtain
\begin{align}
{\cf}_{\zb^{\ab}\yb}&\={\cf}_{\zb^{\ab}\vt}\,\frac{\pa\vt}{\pa\yb}
                       +{\cf}_{\zb^{\ab}\vp}\,\frac{\pa\vp}{\pa\yb}
\=\frac{1}{\yb}\,\left(\sin\vt\,
                       {\cf}_{\zb^{\ab}\vt}-\im\,{\cf}_{\zb^{\ab}\vp}
                       \right)
\=-({\cf}_{z^a y})^\+ \ ,
\nonumber
\\
\label{complfun}
{\cf}_{y\yb}&\={\cf}_{\vt\vp}\,\Bigl|\frac{\pa(\vt,\vp)}{\pa(y,\yb)}\Bigr|
\=\frac{1}{2\im}\,\frac{\sin\vt}{y\yb}\,{\cf}_{\vt\vp}
\=\frac{1}{2\im}\,\frac{(1{+}\cos\vt)^2}{R^2\sin\vt}\,{\cf}_{\vt\vp}\ ,
\end{align}
and one can write the DUY equations on $\rns$
in the alternative form
\begin{equation}\label{D}
2\im\ \delta^{a\bb}\,{\cf}_{z^a\zb^{\bb}}+\sfrac{1}{R^2\sin\vt}
\,{\cf}_{\vt\vp}\=0
\quad,\qquad {\cf}_{\zb^{\ab}\zb^{\bb}}\=0 \quad,\qquad
\sin\vt\, {\cf}_{\zb^{\ab}\vt}-\im\,{\cf}_{\zb^{\ab}\vp}\=0 \ ,
\end{equation}
along with their hermitian conjugates.
It is easy to show that any solution of these $n(n{+}1){+}1$ equations
also satisfies the full Yang-Mills equations.

The transition to the noncommutative DUY equations is trivially achieved
by going over to operator-valued objects everywhere. In particular,
the field strength components in~(\ref{DU1})--(\ref{D}) then read
$\hat{\cf}_{ij}=\pa_{\xh^i}\hat{\ca}_{j}-\pa_{\xh^j}\hat{\ca}_{i}
+[\hat{\ca}_{i},\hat{\ca}_{j}]$,
where $\hat{\ca}_{i}$ are simultaneously $u(2k)$ and operator valued.
To avoid a cluttered notation, we drop the hats from now on.

\section{ Ansatz for the gauge potential }

Rather than attempting to solve the DUY equations (\ref{D}) in full generality,
we shall prescribe a specific $S^2$ dependence for the gauge potential $\ca$,
generalizing an ansatz due to Taubes~\cite{Taubes}. His ansatz\footnote{
Similarly, Witten's ansatz~\cite{Witt} for gauge fields on $\R^4$ 
reduces (\ref{sdym1}) to the vortex equations on the hyperbolic space $H^2$ 
(cf.~\cite{Correa} for noncommutative $\R^4$).} was introduced for SU$(2)$
gauge fields on $\R^2\times S^2$ and reduced the ASDYM equations (\ref{sdym1})
to the vortex equations on $\R^2$. Our ansatz will eliminate
the spherical coordinates and dimensionally reduce the U$(2k)$ DUY equations
on~$\rts$ to generalized coupled vortex equations for a pair of U$(k)$ gauge
fields and a GL$(k,\C)$-valued scalar field living on~$\rt$,
to be described in the next section.

\bigskip

\noindent
{\bf Choice of ansatz.\ }
Our ansatz is motivated by imposing invariance under the SO(3) isometry group
of the two-sphere. Naively, this would seem to lead to the trivial reduction
\begin{equation} \label{Atriv}
{\ca}(x,\vt,\vp) \= {\ca}_\mu(x)~\diff x^\mu \ .
\end{equation}
However, it is natural to allow for gauge transformations
to accompany the SO(3) action~\cite{Forgacs},
and so some ``twisting'' can occur in the reduction.
It is known that the angular dependence in this case is determined via an
embedding of the Dirac monopole (or anti-monopole) U(1)~bundle
$\Lcal\to S^2$ into a trivial SU(2) bundle over
$S^2$~\cite{Taubes}. In the SU(2) Wu-Yang
description~\cite{WuYang}, this monopole is encoded in the matrix
\begin{equation}
Q(\vt,\vp)\ :=\ \im\vec{\s}\cdot\vec{n}(\vt,\vp)
\qquad\textrm{with}\quad \vec{n}^2 =1\ ,
\end{equation}
where $\vec{\s}=(\s_1,\s_2,\s_3)$ denotes the Pauli spin matrices
and $\vec{n}(\vt,\vp)$ parametrizes the unit two-sphere in
three-dimensional space.
In this paper we will consider a more general ansatz by admitting
Dirac {\it multi\/}-monopoles. More concretely, we take $m$ Dirac monopoles
sitting on top of each other ($m\in\Z$) as our angular configuration.\footnote{
We do not turn on the moduli associated with their separation or relative
isospin orientation.}
Note that this is no longer SO(3) invariant for $|m|>1$. Nevertheless,
this generalization is easily performed by choosing~\cite{Bais,Corrigan1}
\begin{equation}\label{Q}
Q\=Q(\vt,m\vp)\=
\im\,(\sin\vt\cos(m\vp)\ \s_1\ +\ \sin\vt\sin(m\vp)\ \s_2\ +\ \cos\vt\ \s_3)
\= -Q^\+ \ ,\quad m\in\Z\ .
\end{equation}

The $su(2)$ valued $m$-monopole connection and curvature read
\cite{Bais,Corrigan1}
\begin{equation}
{\ca}_{su(2)} \= -\sfrac12\,Q~\diff Q \qquad\textrm{and}\qquad
{\cf}_{su(2)} \= -\sfrac14~\diff Q \wedge\diff Q\ .
\end{equation}
The simplest way to embed the $su(2)$ matrix $Q$ into $u(2k)$ consists of the
reduction $u(2k)\to u(k)\otimes\im Q$. This yields
\begin{equation}\label{ourans}
{\ca}(x,\vt,\vp) \= \sfrac12\,A_\mu(x)~\diff x^\mu\otimes\im Q\ +\
\sfrac12\,A_Q(x)\otimes \im Q~\diff Q\ ,
\end{equation}
where $A_\mu$ and $A_Q$ take values in $u(k)$. 
Since noncommutative gauge transformations need the full $u(k)\oplus u(k)$
for closure we should add to this ansatz pieces from $u(k)\otimes\Idd_2$.
Altogether, it implies that our ansatz for the $u(2k)$-valued gauge potential
$\ca$ becomes
\begin{equation}\label{A}
{\ca}(x,\vt,\vp) \= \sfrac{1}{2}\,\bigl\{ A\,\im Q\,+B\,{\Idd}_2\,+\,
(\p_1{-}1)\,Q~\diff Q \,+\, \p_2~\diff Q\bigr\}\ ,
\end{equation}
where the one-forms $A=A_\m(x)~\diff x^\m$ and $B=B_\m(x)~\diff x^\m$
take values in $u(k)$ (i.e. they are anti-hermitian),
while the scalars $\p_{1}=\p_{1}(x)$ and $\p_{2}=\p_{2}(x)$ are
$k\times k$ hermitian matrices.
Note that the fields $(A,B,\p_1,\p_2)$ do not depend on $\vt$ or $\vp$
but act as operators on the Fock space~$\Hcal$ in the noncommutative
gauge theory.

\bigskip

\noindent
{\bf Monopole projectors and gauge potential.\ }
The $su(2)$ matrix $Q$ satisfies the identities
\begin{align}
Q^2&\=-{\Idd}_2 \qquad\Longrightarrow\qquad
(Q~\diff Q)^\+ \= \diff Q~Q \= -Q~\diff Q \ ,
\\
Q\,\frac{\pa Q}{\pa\vp}&\= m\,\sin\vt\,\frac{\pa Q}{\pa\vt}
\qquad\Longrightarrow\qquad
\frac{\pa Q}{\pa\vp}\= -m\,{\sin\vt}\ Q\,\frac{\pa Q}{\pa\vt}\ ,
\label{identities}\\
\diff Q\wedge\diff Q&\= \Bigl( \frac{\pa Q}{\pa\vt}\,\frac{\pa Q}{\pa\vp} -
\frac{\pa Q}{\pa\vp}\,\frac{\pa Q}{\pa\vt}\Bigr)~\diff\vt\wedge\diff\vp
\= -2m\,Q \,\sin\vt~\diff\vt\wedge\diff\vp\ .
\end{align}
It is convenient to introduce the hermitian projectors
\begin{equation} \label{pro}
P_+ \= \sfrac12\,({\Idd}_2 + \im Q) \qquad\textrm{and}\qquad
P_- \= \sfrac12\,({\Idd}_2 - \im Q)
\end{equation}
on $\C^2$ which define the degree $|m|$ monopole and anti-monopole line bundles
$(\Lcal)^m\to S^2$ (with
$(\Lcal)^{-|m|}:=(\,\overline{\Lcal}\,)^{|m|}$). They satisfy
\begin{align}\label{prop}
P_\pm^2&\= P_\pm \quad,\qquad
P_+ + P_- \= {\Idd}_2 \quad,\qquad
P_+ P_- \= 0 \quad,\qquad
\mbox{tr}^{~}_{2\times 2}\,P_{\pm} \= 1\ ,\\[4pt] \label{dP}
\diff P_\pm&\= \Bigl(
\bigl[ (1{-}m)\,P_+ +(1{+}m)\,P_- \bigr]~\e^{\im\vp}~\diff y +
\bigl[ (1{+}m)\,P_+ +(1{-}m)\,P_- \bigr]~\e^{-\im\vp}~\diff \yb \Bigr)
\,\frac{1{+}\cos\vt}{R}\,\frac{\pa P_\pm}{\pa\vt}\ .
\end{align}

The corresponding field combinations are
\begin{equation} \label{A+A-}
A^\pm\ :=\ \sfrac{1}{2}\,(B\pm A)
\qquad\Longleftrightarrow\qquad
A\=A^+ - A^-\ ,\quad B\=A^+ + A^-
\end{equation}
and
\begin{equation} \label{p1p2}
\p\ :=\ \p_1 + \im\,\p_2
\qquad\Longleftrightarrow\qquad
\p_1\=\sfrac{1}{2}\,(\p + \p^\+)\ ,\quad \p_2\=\sfrac{\im}{2}\,(\p^\+ - \p)\ .
\end{equation}
In terms of these degrees of freedom, the ansatz (\ref{A}) becomes
\begin{equation}\label{AA}
{\ca} \= A^+\,P_+\ +\ A^-\,P_- \ +\
(1-\p)\,P_+~\diff P_+\ +\ (1-\p^\+)\,P_-~\diff P_- \ .
\end{equation}
With $A^{\pm}_a:=A^{\pm}_{z^a}$ and $\
A^{\pm}_{\ab}:=A^{\pm}_{\zb^{\ab}}$, this implies
\begin{align}\label{Aa}
{\ca}_a\ :=& \ {\ca}_{z^a}\ =\  A_a^+\,P_+\ +\ A_a^-\,P_-
\= -( {\ca}_{\ab})^\+\ , \\[4pt]
\label{Ay}
\ca_y \ =& \
(1-\p)\,P_+\,\frac{\pa P_+}{\pa y}\ +\ (1-\p^\+)\,P_-\,\frac{\pa P_-}{\pa y}
\ =\ -( {\ca}_{\yb})^\+\ .
\end{align}

\bigskip

\noindent
{\bf Field strength tensor.\ }
The calculation of the curvature
\begin{align}\nonumber
{\cf}\ &=\ \diff {\ca} + {\ca} \wedge {\ca}
\={\sfrac{1}{2}}\,{{\cf}_{ij}}~\diff x^i\wedge\diff x^j \\[4pt]
\label{F}
&=\ \sfrac{1}{2}\,{\cf}_{\m\n}~\diff x^\m\wedge\diff x^\n +
{\cf}_{\m\vt}~\diff x^\m\wedge\diff\vt + {\cf}_{\m\vp}~\diff x^\m\wedge\diff\vp
+ {\cf}_{\vt\vp}~\diff\vt\wedge\diff\vp
\end{align}
for $\ca$ of the form (\ref{A}) yields
\begin{align}\nonumber
4\,{\cf} \ =& \ \bigl( 2~\diff A + A\wedge B + B\wedge A \bigr)\,\im Q\
+\ \bigl( 2~\diff B + A\wedge A + B\wedge B \bigr)\,{\Idd}_2 \\[4pt]
\nonumber
& +\bigl( [\p_1,\p_2]\,Q + (\p_1^2+\p_2^2-1)\,{\Idd}_2 \bigr)~
\diff Q\wedge\diff Q \\[4pt]
\label{FF}
& +\Bigl(\bigl(2~\diff\p_1 +\im A\,\p_2 +\im\p_2\,A +[B,\p_1]\bigr)\,Q
+\bigl(2~\diff\p_2-\im A\,\p_1-\im\p_1\,A+[B,\p_2]\bigr)\,{\Idd}_2
\Bigr)\wedge\diff Q \ .
\end{align}
Rewriting $\cf$ in terms of $A^\pm$ and $\p ,\p^\+$ as in (\ref{AA}), we
obtain
\begin{align}\nonumber
{\cf} \= & F^+\,P_+\ +\ F^-\,P_-\ +\
(1-\p\,\p^\+)\,P_+~\diff P_+\wedge\diff P_+\ +\
(1-\p^\+\,\p)\,P_-~\diff P_-\wedge\diff P_- \\[4pt]
\label{FFF}
& -\ D\phi \wedge P_+~\diff P_+
\ -\ (D\phi)^\+\wedge P_-~\diff P_- \ ,
\end{align}
where $F^\pm:=\diff A^\pm+A^\pm\wedge A^\pm$ and we have introduced
the bi-fundamental covariant derivative
\begin{equation}
D\p \ :=\ \diff\p + A^+\,\p -\p\,A^- \ .
\end{equation}
With $F^{\pm}_{ab}:= F^{\pm}_{z^a z^b}$ and so on, from (\ref{FFF}) we have
\begin{align}\label{Fabbb}
{\cf}_{\ab\bb}\ :=& \ {\cf}_{{\zb}^{\ab}{\zb}^{\bb}}\ =\
F^+_{\ab\bb}\,P_+\ +\ F^-_{\ab\bb}\,P_- \= -( {\cf}_{ab})^\+\ , \\[4pt]
\label{Fabb}
{\cf}_{a\bb}\ :=& \ {\cf}_{z^a{\zb}^{\bb}}\ =\
F^+_{a\bb}\,P_+\ +\ F^-_{a\bb}\,P_- \= -( {\cf}_{\ab b})^\+\ , \\[4pt]
\label{Fyyb}
{\cf}_{y\yb}\ =& \ -\frac{m\,R^2}{(R^2 + y\yb)^2}\
\Bigl\{ (1-\p\,\p^\+)\,P_+\ -\ (1-\p^\+\,\p)\,P_- \Bigr\}
\= -( {\cf}_{\yb y})^\+\ , \\[4pt]
\label{Fzbyb}
{\cf}_{{\zb}^{\ab}{\yb}}\ =& \
-(\bar{D}_{\ab}\p)\,P_+\,\frac{\pa P_+}{\pa\yb}\ -\
(D_{a}\p)^\+ \,P_-\,\frac{\pa P_-}{\pa\yb}
\= -( {\cf}_{z^a y})^\+\ , \\[4pt]
\label{Fzby}
{\cf}_{{\zb}^{\ab}y}\ =& \
-(\bar{D}_{\ab}\p)\,P_+\,\frac{\pa P_+}{\pa y}\ -\
(D_{a}\p)^\+ \,P_-\,\frac{\pa P_-}{\pa y}
\= -({\cf}_{z^a \yb})^\+\ .
\end{align}
Inspecting (\ref{dP}) we notice a simplification for $m=1$, namely
\begin{equation}
\frac{\pa P_\pm}{\pa\yb} \propto P_+ \ ,\quad
\frac{\pa P_\pm}{\pa y}  \propto P_- \qquad\Longrightarrow\qquad
{\cf}_{{\zb}^{\ab}{\yb}} \propto (\bar{D}_{\ab}\p)\,P_+ \ ,\quad
{\cf}_{{\zb}^{\ab} y}    \propto (D_a \p)^\+\,P_- \ .
\end{equation}
For $m=-1$ one interchanges $y$ and $\yb$.
If $|m|\ne1$ then both terms are present in (\ref{Fzbyb}) and
(\ref{Fzby}).

\section{ Generalized coupled vortex equations on $\mbf{\rt}$ }

\noindent
{\bf Reduction to $\mbf{\rt}$.\ }
Let us now see what becomes of the DUY equations on $\rts$
for gauge potentials of the form proposed in the previous section.
Substituting (\ref{Fabbb})--(\ref{Fyyb}) into (\ref{DU1}) and
(\ref{DU2}), now understood in the noncommutative setting, we obtain
\begin{align}\label{dd1}
& \delta^{a\bb}\,F^+_{a{\bb}}\ =\  +\frac{m}{4R^2}\,\left(1-\p\,\p^\+\right)
\qquad\textrm{and}\qquad F^+_{\ab{\bb}}\ =\ 0\= F^+_{ab}\ ,\\ \label{dd2}
& \delta^{a\bb}\,F^-_{a{\bb}}\ =\  -\frac{m}{4R^2}\,\left(1-\p^\+\,\p\right)
\qquad\textrm{and}\qquad F^-_{\ab{\bb}}\ =\ 0\= F^-_{ab}\ .
\end{align}
Setting (\ref{Fzbyb}) to zero (see (\ref{DU3})) leads to
the case distinction
\begin{align}\label{e1}
m=+1 \qquad\Longrightarrow\qquad &
\bar\pa_{\ab}\p + A^+_{\ab}\,\p - \p\,A^-_{\ab}\= 0\\[4pt] \label{e2}
m=-1 \qquad\Longrightarrow\qquad &
\pa_{a}\p + A^+_{a}\,\p - \p\,A^-_{a}\= 0\\[4pt] \label{e1e2}
|m|\ne1 \qquad\Longrightarrow\qquad &
\bar\pa_{\ab}\p + A^+_{\ab}\,\p - \p\,A^-_{\ab} \= 0\qquad{\rm{and}}\qquad
\pa_{a}\p + A^+_{a}\,\p - \p\,A^-_{a} \= 0\ .
\end{align}
We shall call (\ref{dd1})--(\ref{e1e2}) the generalized coupled vortex 
equations.

Instead of working with the gauge potentials $A^\pm_\m$
we shall use the operators $X^{\pm}_\m$ defined via
\begin{equation}\label{X}
X_{a}^{\pm}\ :=\ A_{a}^{\pm} + \th_{a\bb}\,\zb^{\bb}
\qquad\textrm{and}\qquad
X_{{\ab}}^{\pm}\ :=\ A_{{\ab}}^{\pm} + \th_{\ab b}\,z^b\ ,
\end{equation}
in terms of which the field strength tensor reads
\begin{equation}
F_{{a}{\bb}}^{\pm}\ =\ \big[X_{a}^{\pm}\,,\,X_{{\bb}}^{\pm}\big]\ +\ \th_{a\bb}
\qquad\textrm{and}\qquad
F_{{\ab}{\bb}}^{\pm}\ =\ \big[X_{{\ab}}^{\pm}\,,\, X_{\bb}^{\pm}\big]\ .
\end{equation}
Our generalized coupled vortex equations (\ref{dd1})--(\ref{e1e2})
can then be rewritten as
\begin{align}\label{ddd1}
&\delta^{a\bb}\,\Bigl\{\big[X_{a}^{+}\,,\,X_{{\bb}}^{+}\big] +
\th_{a\bb}\Bigr\}
\ =\ +\frac{m}{4R^2}\,\left(1-\p\,\p^\+\right)
\qquad \textrm{and}\qquad
\big[X_{\ab}^{+}\,,\,X_{{\bb}}^{+}\big]\=0\=\big[X_a^{+}\,,\,X_b^{+}\big]\ ,\\
\label{ddd2}
&\delta^{a\bb}\,\Bigl\{\big[X_{a}^{-}\,,\,X_{{\bb}}^{-}\big] +
\th_{a\bb} \Bigr\}
\ =\ -\frac{m}{4R^2}\,\left(1-\p^\+\,\p\right)
\qquad \textrm{and}\qquad
\big[X_{\ab}^{-}\,,\,X_{{\bb}}^{-}\big]\=0\=\big[X_a^{-}\,,\,X_b^{-}
\big]\ ,\\[4pt]
\label{ddd3}
&\qquad m=+1 \qquad\Longrightarrow\qquad
X_{\ab}^{+}\,\p - \p\,X_{\ab}^{-}\ =\ 0\ ,\\[4pt]
\label{ddd4}
&\qquad m=-1 \qquad\Longrightarrow\qquad
X_a^{+}\,\p - \p\,X_a^{-}\ =\ 0\ ,\\[4pt]
\label{ddd5}
&\qquad |m|\ne1 \;\qquad\Longrightarrow\qquad
X_{\ab}^{+}\,\p - \p\,X_{\ab}^{-}\ =\ 0 \qquad{\rm{and}}\qquad
X_a^{+}\,\p - \p\,X_a^{-}\ =\ 0\ .
\end{align}

\bigskip

\noindent
{\bf Features.\ }
The degenerate case $m=0$ corresponds to the trivial reduction (\ref{Atriv})
to~$\rt$, where the vortex potential disappears and the scalars are covariantly
constant in the background of a DUY gauge potential on~$\rt$.
One must also distinguish between $|m|=1$ and $|m|>1$ because
these two cases are quite different in nature.
For $m=\pm\,1$ the gauge potential~$\ca$ is spherically symmetric
(up to a gauge transformation). Eqs.~(\ref{ddd3}) and (\ref{ddd4}) are
related by the interchanges $z^a\leftrightarrow{\zb}^{\ab}$ and
$y\leftrightarrow\yb$ ($\vp\leftrightarrow -\vp$).
The field strength component ${\cf}_{\vt\vp}$ is
proportional to the angular part of the Dirac monopole
(or anti-monopole) field strength tensor. One may regard (\ref{e1})
and (\ref{e2}) as, respectively, a `holomorphicity' and
`anti-holomorphicity' condition for~$\p$. The commutative analogue of
(\ref{dd1})--(\ref{e2}) was studied in the mathematical literature
(see e.g.~\cite{Prada,BGP}) under the name of `coupled vortex equations'.
More generally, the situation $|m|>1$ corresponds to $m$ Dirac (anti-)monopoles
sitting on top of each other.
These configurations are not spherically symmetric.
Furthermore, the requirement of `covariant constancy'
(\ref{e1e2}) implies the compatibility conditions
\begin{equation}\label{e3}
F^+_{\m\n}\,\p -\p\,F^-_{\m\n}\=0
\end{equation}
which are rather restrictive.
Nevertheless, we shall see that they can be satisfied non-trivially for
the noncommutative instanton configurations to be considered later on.

Finally, we remark that for $n=2$, $m=1$ and $k=1$ the equations
(\ref{dd1})--(\ref{e1}) coincide with the perturbed Seiberg-Witten
${\rm U}_+(1)\times {\rm U}_-(1)$ monopole equations on $\R^4_\th$
as considered in~\cite{PSW}. In the commutative limit they reduce to
the standard perturbed abelian Seiberg-Witten monopole equations on $\R^4$.
The perturbation, i.e. the term $\frac{1}{4R^2}$ in (\ref{dd1}) and
(\ref{dd2}), is introduced into the Seiberg-Witten equations by hand.
In the present context, it arises automatically from
the extra space $S^2$ and the reduction from $\R^4\times S^2$ to~$\R^4$.

\section{ BPS solutions of the Yang-Mills equations on $\mbf{\rts}$ }

We are now ready to construct solutions to the generalized coupled vortex
equations (\ref{ddd1})--(\ref{ddd5}), and thus to the DUY equations on~$\rts$,
by making use of partial isometries and the (noncommutative) ABS
construction.

\bigskip

\noindent
{\bf Explicit solutions.\ }
Let us consider the ansatz
\begin{align}\label{ansatz1}
X_{a}^{\pm} \ &=\ \th_{a\bb}\,T^{~}_{N_{\pm}}\,\zb^{\bb}\,T_{N_{\pm}}^\+
\qquad\Longrightarrow\qquad
X_{{\ab}}^{\pm} \=\th_{\ab b}\, T^{~}_{N_{\pm}}
\,z^{b}\,T_{N_{\pm}}^\+ \ ,\\[4pt]
\label{ansatz2}
\p \ &=\ T^{~}_{N_{+}}T_{N_{-}}^\+\qquad\qquad\ \Longrightarrow\qquad\quad
\p^\+ \ = \ T^{~}_{N_{-}}T_{N_{+}}^\+\ ,
\end{align}
where $T^{~}_{N_{\pm}}$ are $k\times k$ matrices (with operator entries
acting on $\Hcal$) possessing the properties
\begin{align}\label{T1}
T_{N_{+}}^\+ T^{~}_{N_{+}}\=1\qquad\textrm{while}\qquad
T^{~}_{N_{+}}T_{N_{+}}^\+\ =\ 1 - P^{~}_{N_{+}}\ , \\
\label{T2}
T_{N_{-}}^\+ T^{~}_{N_{-}}\=1\qquad\textrm{while}\qquad
T^{~}_{N_{-}}T_{N_{-}}^\+\ =\ 1 - P^{~}_{N_{-}}\ .
\end{align}
Here $P^{~}_{N_{+}}$ and $P^{~}_{N_{-}}$ are projectors of rank $N_{+}$ and
$N_{-}$,
respectively, on the Fock space ${\mathbb C}^k\otimes{\Hcal}$, where $\Hcal$
denotes the $n$-oscillator Fock space. In other words, the operators
$T^{~}_{N_\pm}$ are partial isometries on $\C^k\otimes\Hcal$.
This ansatz directly yields
\begin{align}\label{ansF}
&F_{a\bb}^{\pm} \= \th_{a\bb}\,P^{~}_{N_{\pm}} \=
\frac{1}{2\th^a}\,\de_{a\bb}\,P^{~}_{N_{\pm}}
\qquad\textrm{and}\qquad
F_{{\ab}{\bb}}^{\pm} \=0\= F_{ab}^{\pm}\ ,\\[4pt]
\label{ansp}
&1-\p\,\p^\+ \= P^{~}_{N_+} \qquad\textrm{and}\qquad 1-\p^\+\,\p \=
P^{~}_{N_-}\ .
\end{align}

Inserting these expressions into our generalized coupled vortex equations
(\ref{ddd1})--(\ref{ddd5}), we find that (\ref{ddd3})--(\ref{ddd5})
are automatically satisfied. The first two equations reduce to the
constraints\footnote{
It is possible to generalize the ansatz (\ref{ansatz1}) and~(\ref{ansatz2}),
following the discussion of flux tubes and vortices in two dimensions 
\cite{Poly, Bak}. In this way, the constraints (\ref{res}) and~(\ref{res-}) 
may be relaxed~\cite{Bak}.}
\begin{align}\label{res}
(\ref{ddd1})\quad\Longrightarrow\qquad
\frac{1}{\th^1}+\ldots + \frac{1}{\th^n} &\= +\frac{m}{2R^2}
\qquad\textrm{if}\quad N_+\ne0\ ,\\
\label{res-}
(\ref{ddd2})\quad\Longrightarrow\qquad
\frac{1}{\th^1}+\ldots + \frac{1}{\th^n} &\= -\frac{m}{2R^2}
\qquad\textrm{if}\quad N_-\ne0\ .
\end{align}
For any $m$ and $\th^a>0$ (which we assume) 
these two conditions are incompatible,
implying that our ansatz does not allow for BPS configurations with both
$N_+\ne0$ and $N_-\ne0$. However, we shall show in the next section that
in this case the configuration (\ref{ansatz1}), (\ref{ansatz2}) still
satisfies the {\it full\/} Yang-Mills equations on $\rts$!
Yet, in order to solve (\ref{ddd1}) and (\ref{ddd2}) simultaneously one must
put either $N_+=0$ or $N_-=0$. For example, if $N_-=0$ we have
$T^{~}_{N_-}=1$ and thus
\begin{equation}
X_a^-=\th_{a\bb}\,\zb^{\bb} \quad\Longrightarrow\quad
A_a^-=0=A_{\ab}^- \quad\Longrightarrow\quad
F^-=0 \qquad\textrm{and}\qquad
1-\p^\+\,\p = 1-T^{~}_{N_{-}}T_{N_{-}}^\+=0\ ,
\end{equation}
so that (\ref{ddd2}) becomes an identity and only (\ref{res}) remains.
Likewise, $N_+=0$ gives viable BPS configurations. Putting $N_+=0=N_-$,
however, yields only the vacuum solution $(A{=}0,\p{=}1)$.
Note also that the case $m=0$ is in conflict with the positivity of
all~$\th^a$.

These observations have a natural physical interpretation.
The original DUY equations are fixed by the parameters $n$ and~$k$.
Our ansatz (\ref{AA}) together with (\ref{ansatz1}), (\ref{ansatz2})
is labelled by the triple $(m,N_+,N_-)$. According to the standard
identifications of D-branes as noncommutative solitons~\cite{Strom},
for $|m|=1$ it should describe  a collection of $N_+$ D$0$-branes and $N_-$
anti-D$0$-branes as a bound state (i.e. a vortex-like solution on
$\R^{2n}_\th$) in a system of $k$ ${\rm D}(2n)-\overline{{\rm D}(2n)}$
brane-antibrane pairs. It is known that such a bound state can only be
stable (i.e. possess the BPS property) if either $N_+=0$ or $N_-=0$,
which fits perfectly with our findings. Thus, the DUY configurations
with $|m|=1$ obtained from (\ref{ansatz1}), (\ref{ansatz2}) are stable 
bound states of ${\rm  D}(2n)-\overline{{\rm D}(2n)}$ pairs containing 
either $N_+$ D$0$-branes or $N_-$ anti-D$0$-branes, but not both. The 
D-brane interpretation of our solutions will be elucidated in more detail
later on.

\bigskip

\noindent
{\bf Explicit realization of the operators $\mbf{T^{~}_{N_{\pm}}}$.\ }
For the operators $T^{~}_{N_{\pm}}$ one may take the matrices
$T^{~}_N$ from~\cite{PSW} (and references therein). Namely,
\begin{equation}\label{TT}
\quad T^{~}_N\=(T)^N \qquad \mbox{with}\qquad T^\+ \=
\frac{1}{\sqrt{(\g \cdot x)(\g \cdot x)^\+}}\,\g \cdot x\ ,
\end{equation}
where $\g \cdot x := \g_\mu\,x^{\mu}$, and the $k\times k$ matrices
${\g}_{\mu}$ are subject to the anti-commutation relations\footnote{
In this part we are more explicit regarding the matrix structure.}
\begin{equation}
{\g_\mu}^\+\,\g_\nu + {\g_\nu}^\+\,\g_\mu \= 2\,\de_{\mu\nu}\,\Idd_k \=
\g_\mu\,{\g_\nu}^\+ + \g_\nu\,{\g_\mu}^\+ \ .
\end{equation}
This implies that
\begin{equation}\label{G}
\Gamma_\mu\ = \ \begin{pmatrix}0&{{\g_\mu}^\+}\,\\ -\g_\mu & 0\end{pmatrix}
\qquad\textrm{satisfies}\qquad
\Gamma_\mu\,\Gamma_\nu + \Gamma_\nu\,\Gamma_\mu \= -2\,
\de_{\mu\nu}\,\Idd_{2k}\ ,
\end{equation}
i.e.~they generate the Clifford algebra $\cliff_{2n}$ of the inner
product space $(\R^{2n}, \de_{\mu\nu})$. Hence the choice (\ref{TT})
restricts us to $k=2^{n-1}$. Note that for $n=1\ \Leftrightarrow\ k=1$
we have $\gamma_1=1$, $\gamma_2=\im$, which yields
\begin{equation}
T^\+\=\frac{1}{\sqrt{\zb z}}\,\zb \= \sum_{k=1}^\infty |k{-}1\>\<k|\ ,
\end{equation}
and we obtain the standard shift operator $S_N=(T)^N$ in this case.

In general, the operator (\ref{TT}) may be regarded as a map
\begin{equation}
T^{~}_N\,:\ \Delta^-\otimes\Hcal~\longrightarrow~\Delta^+\otimes\Hcal \ ,
\label{TTop}\end{equation}
where $\Delta^\pm\cong\C^k$ are the irreducible chiral spinor modules of
dimension $k=2^{n-1}$ on which the matrices $\gamma_\mu$ act.
It is not difficult to see that
\begin{equation}\label{TN}
T_N^\+ T^{~}_N \= \Id\otimes1 \qquad\textrm{while}\qquad
T^{~}_N T_N^\+\= \Id\otimes1 - P^{~}_N\ ,
\end{equation}
where $P^{~}_N$ is a rank-$N$ projector on the space ${\mathbb
  C}^k\otimes\Hcal$. In particular, the operator $T=T_1$ has no
kernel, while $T^\dag$ has a one-dimensional kernel which is spanned
  by the vector $|\alpha\>\otimes|0,\ldots,0\>$ where $|\alpha\>$
  denotes the lowest-weight spinor of SO($2n$). On the other hand,
  $T^\dag$ is surjective, ${\rm
  im}(T^\dag)=\Delta^-\otimes\Hcal$. Consequently,
\begin{equation}
\dim\ker(T^{~}_N)=0 \qquad\textrm{but}\qquad\dim\,{\rm coker}(T^{~}_N)=N \ .
\label{dimkerTN}\end{equation}
Operators satisfying such conditions are known as Toeplitz operators
and generate an algebra called the Toeplitz algebra. The construction
(\ref{TT}) is known as the (noncommutative) ABS
construction~\cite{HM1}. It provides a convenient realization of the
operators $T^{~}_{N_\pm}$ which we will exploit later on when we
analyze explicitly the brane interpretation of our noncommutative
multi-instanton solutions. However, even there, only the generic
properties of these operators are really important.

\bigskip

\noindent
{\bf Generic form of $\mbf{T^{~}_N}$.\ }
The realization (\ref{TT}) can be generalized in order to introduce
$2nN$ real moduli into the solution, specifying the locations of the
$N$ noncommutative solitons in $\R^{2n}$. For this, we will write
$T^{~}_N=U_1\,U_2\cdots U_N$  where each $U_\ell^\+$ is of the form of
$T^\+$ in (\ref{TT}) but with  the coordinates~$x$ shifted to
$x_\ell:=x-b_\ell$.
Let us illustrate this strategy on the example of $n=2\ \Leftrightarrow\ k=2$
for the special case of equal noncommutativity parameters $\th^1=\th^2$.
We redenote the complex coordinates $(z^1,z^2)$ on $\R^4_\th$ by $(y,z)$ and
represent the $2\times 2$ matrices $T$ and $T^\+$ of (\ref{TT}) as\footnote{
These matrices can be copied from \cite{LP2} and references therein.}
\begin{equation}\label{UU}
U   \= \begin{pmatrix}z&y\\ \yb&-\zb\end{pmatrix}\,\frac{\im}{r}
\qquad\textrm{and}\qquad
U^\+\=-\frac{\im}{r}\,\begin{pmatrix}\zb&y\\ \yb&-z\end{pmatrix}
\qquad\textrm{with}\quad r\=\sqrt{y\yb +\zb z}\ .
\end{equation}
It is easily checked that they fulfill
\begin{equation}
U^\+\,U \= \begin{pmatrix}1&0\\0&1\end{pmatrix}\qquad\textrm{while}\qquad
U\,U^\+ \= \begin{pmatrix}1-|0,0\>\<0,0|&0\\0&1\end{pmatrix}
\end{equation}
and hence the kernel of $U^\+$ is spanned by the vector
${1\choose0}\otimes|0,0\>$. We now introduce the shifted matrices
\begin{equation}\label{Ul}
U_\ell^\+\ =\ -\frac{\im}{r_\ell}\,
\begin{pmatrix}\zb_\ell&y_\ell\\ \yb_\ell & -z_\ell\end{pmatrix}
\qquad\textrm{with}\quad
y_\ell :=y-b^y_\ell\ ,\quad z_\ell:=z-b^z_\ell\ ,\quad \ell=1,\ldots,N
\end{equation}
and $r_\ell=\sqrt{y_\ell \yb_\ell + \zb_\ell z_\ell}$.
They behave just like $U^\+$ in (\ref{UU}) except that the kernel is
modified according to
\begin{equation}
U_\ell^\+\ {\textstyle{1\choose0}}\otimes|{\bb}_\ell\> \=0
\qquad\textrm{where}\qquad
\yb_\ell|{\bb}_\ell\>\=0\=\zb_\ell|{\bb}_\ell\>\ ,
\end{equation}
i.e.~$|{\bb}_\ell\>$ is a coherent state depending on the two complex
parameters $b_\ell^y$ and~$b_\ell^z$.

Consider then the states
\begin{equation}\label{xil}
|\!|\xi_1\>\!\> \ :=\ {\textstyle{1\choose0}}\otimes|{\bb}_1\>
\qquad\textrm{and}\qquad
|\!|\xi_\ell\>\!\> \ :=\ U_1\cdots U_{\ell-1}\,
{\textstyle{1\choose0}}\otimes|{\bb}_\ell\>
\qquad\textrm{for}\quad \ell=2,\ldots,N\ .
\end{equation}
Clearly, they are all annihilated by the operator
\begin{equation}\label{TN+}
T_N^\+\ :=\ U^\+_N\cdots U^\+_\ell \cdots U^\+_1
\end{equation}
which indeed obeys
\begin{equation}
T_N^\+\,T^{~}_N \= \Idd_2\otimes1 \qquad\textrm{while}\qquad
T^{~}_N\,T_N^\+ \= \Idd_2\otimes1 - P^{~}_N\ ,
\end{equation}
where $P^{~}_N$ is the orthogonal projection onto the $N$-dimensional
subspace in  ${\mathbb C}^2\otimes\Hcal$ spanned by the vectors
$|\!|\xi_1\>\!\>,\ldots,|\!|\xi_N\>\!\>$.  Similarly, one can
introduce operators $T^{~}_N$ for $n>2$ generalizing (\ref{TT}).

\section{ Non-BPS solutions of the Yang-Mills equations on
  $\mbf{\rts}$ }

The DUY equations on $\rts$ are BPS conditions for the
Yang-Mills equations on $\rts$. Therefore, the configuration
(\ref{ansatz1}), (\ref{ansatz2}) with $T^{~}_{N_+}=1$ or with $T^{~}_{N_-}=1$
produces BPS solutions of the Yang-Mills equations on $\rts$.
We saw that configurations with both $N_+$ and $N_-$ being non-zero
did not satisfy the DUY equations. We shall now demonstrate that
for any value of $m$, $N_+$ and $N_-$ these still yield non-BPS solutions
of the full Yang-Mills equations, with topological charge ${\cal
  Q}=m\,(N_+{-}N_-)$.

We recall that
\begin{align}
\label{chia}
\ca_a- \th_{a\bb}\,\zb^{\bb} &\=
X^+_a\,P_+ + X^-_a\,P_-\ ,\\[4pt]
\label{chiab}
\ca_{\ab} - \th_{\ab b}\, z^{b} &\=
X^+_{\ab}\,P_+ + X^-_{\ab}\,P_-\ ,\\
\label{avt}
\ca_\vt &\=
(1-\p)\,P_+\,\frac{\pa P_+}{\pa\vt} + (1-\p^\+)\,P_-
\,\frac{\pa P_-}{\pa\vt}\ ,\\
\label{avp}
\ca_\vp &\=
(1-\p)\,P_+\,\frac{\pa P_+}{\pa\vp} + (1-\p^\+)\,P_-\,\frac{\pa P_-}{\pa\vp}\ .
\end{align}
For the ansatz (\ref{ansatz1}) and (\ref{ansatz2}), $X^{\pm}_{\m}$ and $\p$
are expressed in terms of $T^{~}_{N_\pm}$, and we get
\begin{align}
\label{cfabb}
\cf_{a\bb} &\= \th_{a\bb}\, \{P^{~}_{N_+}\,P_+ + P^{~}_{N_-}\,P_- \}\ , \\[4pt]
\label{cfvtvp}
\cf_{\vt\vp} &\= -\im\,\sfrac{m}{2}\,
\sin\vt\, \{P^{~}_{N_+}\,P_+ - P^{~}_{N_-}\,P_- \}\ ,
\end{align}
with all other components of ${\cf}_{ij}$ vanishing. Let us now insert
these expressions into the Yang-Mills equations which have the form
\begin{equation}\label{YMM}
\frac{1}{\sqrt{g}}\, \pa_i(\sqrt{g}~\cf^{ij}) + [\ca_i, \cf^{ij}]=0
\qquad\Longrightarrow\qquad
\pa_i(\sqrt{g}~g^{ik}\,g^{jl}\,\cf_{kl}) +
\sqrt{g}~g^{ik}\,g^{jl}\,[\ca_i, \cf_{kl}]=0\ ,
\end{equation}
where $\sqrt{g}:=\sqrt{\det(g_{ij})}= R^2\sin\vt$.
It is enough to consider the cases $j=c$ and $j=\vt ,\vp$ since the
case $j=\cb$ can be obtained by hermitian conjugation of (\ref{YMM})
due to the anti-hermiticity of $\ca_i$ and~$\cf_{ij}$.

For $j=c$, (\ref{YMM}) reduces to
\begin{equation}\label{rYM}
g^{\cb a}\,g^{\bb
c}\,\bigl(\pa_{\cb}\cf_{a\bb}+[\ca_{\cb},{\cf}_{a\bb}]\bigr)\=0
\qquad\Longleftrightarrow\qquad
g^{\cb a}\,g^{\bb c}\,[\ca_{\cb}-\th_{\cb b}\,z^{b},{\cf}_{a\bb}]\=0\ .
\end{equation}
Substituting (\ref{chiab}) and (\ref{cfabb}),
we see that (\ref{rYM}) is satisfied due to the identities (\ref{prop}),
(\ref{T1}) and (\ref{T2}). In the case $j=\vt$, (\ref{YMM}) simplifies to
\begin{equation}\label{rrYM}
\pa_\vp (\sqrt{g}~g^{\vt\vt}\,g^{\vp\vp}\,{\cf}_{\vt\vp}) + \sqrt{g}~
g^{\vt\vt}\,g^{\vp\vp}\,[\ca_\vp ,\cf_{\vt\vp}] \=0
\end{equation}
which turns out to be satisfied identically. Likewise, for $j=\vp$ one obtains
\begin{equation}\label{rrrYM}
\pa_\vt (\sqrt{g}~g^{\vt\vt}\,g^{\vp\vp}\,{\cf}_{\vt\vp}) + \sqrt{g}~
g^{\vt\vt}\,g^{\vp\vp}\,[\ca_\vt ,\cf_{\vt\vp}] \=0
\end{equation}
which is true as well.
Hence, the Yang-Mills equations on $\rts$ are satisfied, and we have found
non-BPS configurations of the form (\ref{ansatz1}), (\ref{ansatz2})
where both $N_+$ and $N_-$ are non-zero.

\section{ The topological charge }

Let us now compute the topological charge (the $(n{+}1)$-th Chern 
number) of the above configurations.
This calculation is similar to the one in~\cite{IL}. Namely,
\begin{align}
{\cf}_{2a-1\, 2a} \= &2\im\, {\cf}_{a\ab} \= -\sfrac{\im}{\th^a}\,
\bigl\{ P^{~}_{N_+}\,P_+ + P^{~}_{N_-}\,P_- \bigr\} \ ,\\[4pt]
{\cf}_{\vt\vp} \= &-\im\,\sfrac{m}{2}\,\sin\vt\,
\bigl\{P^{~}_{N_+}\,P_+ - P^{~}_{N_-}\,P_-\bigr\} \\[4pt]
\Longrightarrow\quad
{\cf}_{12}\,{\cf}_{34}\cdots{\cf}_{2n-1\, 2n}\,{\cf}_{\vt\vp}\=
&(-\im)^{n+1}\,\frac{m\,\sin\vt}{2\,\prod\limits^n_{a=1}\th^a}\,
\bigl\{ P^{~}_{N_+}\,P_+ + P^{~}_{N_-}\,P_- \bigr\}^n \,
\bigl\{ P^{~}_{N_+}\,P_+ - P^{~}_{N_-}\,P_- \bigr\}\nonumber\\ \=&
(-\im)^{n+1}\,\frac{m\,\sin\vt}{2\,\prod\limits^n_{a=1}\th^a}\,
\bigl\{ P^{~}_{N_+}\,P_+ - P^{~}_{N_-}\,P_- \bigr\} \\ \label{tr}
\Longrightarrow\quad\mbox{tr}^{~}_{2\times 2}
({\cf}_{12}\,{\cf}_{34}\cdots{\cf}_{2n-1\, 2n}\,{\cf}_{\vt\vp}) &\=
(-\im)^{n+1}\,\frac{m\,\sin\vt}{2\prod\limits^n_{a=1}\th^a}\,
\bigl\{P^{~}_{N_+} - P^{~}_{N_-}\bigr\}\ ,
\end{align}
where we have used the identities (\ref{prop}). It follows that
\begin{align}
\mbox{tr}^{~}_{2\times 2}\,
\underbrace{{\cf}\wedge\ldots\wedge {\cf}}_{n+1} &\=
(n{+}1)!\ \mbox{tr}^{~}_{2\times 2}\,
{\cf}_{12}\,{\cf}_{34}\cdots{\cf}_{2n-1\,2n}\,{\cf}_{\vt\vp}\
\diff x^1\wedge\diff{x^2}\wedge\ldots\wedge\diff x^{2n}
\wedge\diff\vt\wedge\diff\vp \nonumber\\
&\=(n{+}1)!\,(-\im)^{n+1}\,\frac{m\,(P^{~}_{N_+}{-}P^{~}_{N_-})}
{2\,\prod\limits_{a=1}^n{\th^a}}~
\diff x^1\wedge\diff{x^2}\wedge\ldots\wedge\diff x^{2n}
\wedge\sin\vt~\diff\vt\wedge\diff\vp\ .
\end{align}
With this, the topological charge indeed becomes
\begin{align}
{\cal Q}\ :=\ &\frac{1}{(n{+}1)!}\ \Bigl(\frac{\im}{2\pi}\Bigr)^{n+1}\,
\Bigl(\prod_{a=1}^n{2\pi\th^a}\Bigr)~\mbox{Tr}^{~}_{\C^k\otimes\cal H}\,
\int_{S^2} \mbox{tr}^{~}_{2\times 2}\,
\underbrace{{\cf}\wedge\ldots\wedge {\cf}}_{n+1} \nonumber\\
\= &\Bigl(\frac{\im}{2\pi}\Bigr)^{n+1}\,(-\im)^{n+1}\,\frac{m}{2}\,
\Bigl(\prod_{a=1}^n{2\pi\th^a}\Bigr) \Bigl(\mbox{Tr}^{~}_{\C^k\otimes\cal H}
\,\frac{(P^{~}_{N_+}{-}P^{~}_{N_-})}{\prod\limits_{b=1}^n{\th^b}}\Bigr)\
\int_{S^2} \sin\vt~\diff\vt\wedge\diff\vp \nonumber\\
\= &\frac{m}{4\pi}\ \mbox{Tr}^{~}_{\C^k\otimes\cal H}
(P^{~}_{N_+} - P^{~}_{N_-})\
\int_{S^2} \sin\vt~\diff\vt\wedge\diff\vp \nonumber\\[4pt]
\=& m\ {\mbox{Tr}}^{~}_{\C^k\otimes\cal H} (P^{~}_{N_+} - P^{~}_{N_-})
\= m\ ({N_+} - {N_-})\ .
\label{TopchargeYM}\end{align}

\section{ The Yang-Mills functional }

Besides the topological charge, it is also instructive to know the
value of the (Euclidean) action functional on our solutions.
For U($2k$) Yang-Mills theory on $\rts$ this functional has the form
\begin{equation} \label{action}
E \= - \frac1{4g_{\rm YM}^2}\,\Bigl(\prod_{a=1}^n{2\pi\th^a}\Bigr)\
\mbox{Tr}^{~}_{\cal H}\,
\int_{S^2} \diff\vt~\diff\vp~R^2 \sin\vt\ \mbox{tr}^{~}_{2k\times 2k}
\bigl\{ {\cf}_{ij}\,{\cf}^{ij} \bigr\} \ ,
\end{equation}
with $g^{~}_{\rm YM}$ the Yang-Mills coupling constant.
This Euclidean action may be interpreted as an energy functional for
static Yang-Mills fields in $(2n{+}2)+1$ dimensions.

For our ansatz (\ref{A}) with a fixed dependence on the spherical coordinates
$\vt$ and~$\vp$, this functional can be reduced to an integral
($\mbox{Tr}^{~}_{\C^k\otimes\cal H}$) over $2n$ dimensions. Substituting
(\ref{Fabbb})--(\ref{Fzby}) into (\ref{action}), and performing the
integral over $S^2$ and the trace over $\C^2$ in $\C^{2k}=\C^2\otimes\C^k$,
we arrive at
\begin{align} \nonumber
E \= \frac{2\pi R^2}{g_{\rm YM}^2}\, \Bigl(\prod_{a=1}^n{2\pi\th^a}\Bigr)\
\mbox{Tr}^{~}_{\C^k\otimes\cal H} \Bigl\{ &|F^+|^2\ +\ |F^-|^2\ +\
\sfrac{m^2+1}{4R^2}\,|D\p|^2 + \sfrac{m^2+1}{4R^2}\,|(D\p )^\+|^2 \\
\label{energy}
&+\ \sfrac{m^2}{4R^4}\,(1-\p\,\p^\+)^2\
 +\ \sfrac{m^2}{4R^4}\,(1-\p^\+\,\p)^2\Bigr\}\ ,
\end{align}
where we have introduced the customary shorthand notation~\cite{Jaffe}
\begin{equation}
|F|^2 \ :=\ \sfrac12\,{F_{\mu\nu}}^\+\,F^{\mu\nu} \qquad\textrm{and}\qquad
|D\p|^2 \ :=\ (D_\mu \p)^\+\,(D^\mu \p) \ .
\end{equation}
Note that the relative normalization of the $|F|^2$ and $|D\p|^2$ terms
is not important since it can be changed by rescaling the coordinates
\cite{Jaffe}.
By a Bogomolny type transformation one can show that
solutions to the generalized coupled vortex equations
(\ref{dd1})--(\ref{e1e2}) for $m$ fixed realize absolute minima of
the action functional~(\ref{energy}).\footnote{
For a discussion in the undeformed case see~\cite{Prada}.}

Let us evaluate the action functional (\ref{energy}) on our solutions
(\ref{ansatz1})--(\ref{ansp}) to the Yang-Mills equations on~$\rts$,
first without assuming the BPS property (\ref{res}) or (\ref{res-}).
Recall that our ansatz (\ref{ansatz1}), (\ref{ansatz2}) automatically fulfills
\begin{equation}
D_\mu\p\=0 \qquad\textrm{and}\qquad
1-\p\,\p^\+ \= P^{~}_{N_+}\ ,\quad 1-\p^\+\,\p \= P^{~}_{N_-}\ .
\end{equation}
If we assume again for simplicity that $\th^1=\ldots=\th^n=:\th$ then we get
\begin{equation}
|F^+|^2 + |F^-|^2 \= 4\,\de^{a\cb}\,\de^{d\bb}\,
\bigl\{ F^+_{a\bb}\,F^+_{d\cb} + F^-_{a\bb}\,F^-_{d\cb} \bigr\} \=
\frac{n}{\th^2}\,\bigl\{ P^{~}_{N_+} + P^{~}_{N_-} \bigr\}\ .
\end{equation}
Substituting these expressions into (\ref{energy}), we finally obtain
\begin{equation}
E \= \frac{2\pi R^2}{g_{\rm YM}^2}
\,(2\pi\th)^n\,\Bigl(\frac{n}{\th^2}+\frac{m^2}{4R^4}\Bigr)\,
(N_+ + N_-)\ .
\label{energyinst}\end{equation}
The first term yields the tension appropriate to $(N_++N_-)$ D0-branes
on a D$(2n)$-brane~\cite{Strom} in the Seiberg-Witten decoupling
limit, times the area $4\pi R^2$ of the auxiliary sphere $S^2$. The
second term is proportional to the Yang-Mills energy on $S^2$ of $|m|$
coincident Dirac monopoles. In the BPS case ($N_+{=}0$ or $N_-{=}0$)
we can use the relation
\begin{equation}
|m|\,\th \= 2\,n\,R^2
\end{equation}
{}from (\ref{res}) or (\ref{res-}) to write (\ref{energyinst}) as
\begin{equation}
E_{\rm BPS} \= \frac1{2g_{\rm YM}^2}\,
(2\pi)^{n+1}\,\th^{n-1}\,(n{+}1)\,|m|\,N_{\pm}\ .
\label{energyBPS}\end{equation}

\section{ Special solutions }

The extremal cases $\phi =0$ and $\phi =1$ are worth special
consideration. We shall now study them in some detail.

\bigskip

\noindent
{\bf Zero monopole sector.\ }
Let us first look at the case $\phi =\phi^\+ =1$.
The generalized coupled vortex equations (\ref{dd1})--(\ref{e1e2}) then imply
$A^+=A^-=:A$ and $F^+=F^-=:F$ as well as
\begin{equation} \label{phi0F}
\de^{a\bb}\,F_{a\bb} \= 0 \qquad\textrm{and}\qquad
F_{\ab\bb} \= 0 \= F_{ab}\ ,
\end{equation}
which are simply the DUY equations on $\rt$
(note that (\ref{Fyyb}) implies ${\cf}_{y\yb}=0={\cf}_{\vt\vp}$ if $\p=1$).
This is equivalent to putting $m=0$, i.e. the trivial reduction~(\ref{Atriv}).
Inside the ansatz (\ref{ansatz1}), (\ref{ansatz2}) we get\footnote{
However, one may consider (\ref{ansatz1}) with $N_+=N_-\ne0$ without
(\ref{ansatz2}) and obtain non-trivial solutions on $\rt$ after relaxing the 
condition $\th^a > 0$ for all $a$ and allowing at least one of the $\th^a$ to
be negative.}
\begin{equation}
N_+ \= N_- \= 0 \qquad\Longrightarrow\qquad P^{~}_{N_\pm} \= 0
\qquad\Longrightarrow\qquad F^{\pm} \= 0 \ .
\end{equation}
This sector can be understood physically as the endpoint of tachyon
condensation, wherein the tachyon field $\phi$ has rolled to its
minimum at $\phi=\phi^\dag=1$ and all the flux has been radiated away
to infinity. Here the D0-branes have been completely dissolved into
the D$(2n)$-branes.

\bigskip

\noindent
{\bf Zero tachyon sector.\ }
The choice $\phi =0$ is more interesting since from (\ref{Fyyb}) we
then have ${\cf}_{\vt\vp}=-\im\,\frac{m}{2}\,\sin\vt\,\{P_+{-}P_-\}$.
{}For this case (\ref{dd1})--(\ref{e1e2}) reduce to
\begin{align}\label{rduy1}
&\de^{a\bb}\,F^+_{a\bb} \= +\frac{m}{4R^2} \qquad\textrm{and}\qquad
F^+_{\ab\bb} \= 0 \= F^+_{ab}\ ,\\
\label{rduy2}
&\de^{a\bb}\,F^-_{a\bb} \= -\frac{m}{4R^2} \qquad\textrm{and}\qquad
F^-_{\ab\bb} \= 0 \= F^-_{ab}\ .
\end{align}
At $m{=}0$ this includes the previous case of $\phi=1$.
After switching to the matrix form via (\ref{X}) we obtain
\begin{align}\label{mf1}
&\de^{a\bb}\,\big[X^+_{a}\,,\,X^+_{\bb}\big] + \de^{a\bb}\,
\th_{a\bb} - \frac{m}{4R^2}\=0
\qquad\textrm{and}\qquad \big[X^+_{\ab}\,,\,X^+_{\bb}\big]\=0\=
\big[X^+_a\,,\,X^+_b\big] \ , \\
\label{mf2}
&\de^{a\bb}\,\big[X^-_{a}\,,\,X^-_{\bb}\big] + \de^{a\bb}\,
\th_{a\bb} + \frac{m}{4R^2}\=0
\qquad\textrm{and}\qquad \big[X^-_{\ab}\,,\,X^-_{\bb}\big]\=0\=
\big[X^-_a\,,\,X^-_b\big] \ .
\end{align}
Since $\p=0$ cannot be reached from (\ref{ansatz1}), (\ref{ansatz2}),
we have to look outside this ansatz for solving (\ref{mf1}) and
(\ref{mf2}). It gives the local maximum of the tachyon potential
corresponding to the open string vacuum containing D-branes.

\bigskip

\noindent
{\bf Explicit solutions with $\mbf{\p=0}$.\ }
Let us restrict ourselves to the abelian case $k=1$ 
and simplify matters by taking $\th^a=\th$ for all $a=1,\ldots,n$.
We consider the alternative ansatz~\cite{Kraus,Nekrasov}
\begin{equation}\label{ans}
X^\pm_a\=\th_{a\cb}\, S_{l_\pm}^\+\,f_{\pm}(\Ncal\,)\, \zb^{\cb}\,S_{l_\pm}
\qquad\textrm{and}\qquad
X^\pm_{\ab}\=\th_{\ab d}\, S_{l_\pm}^\+\,z^d\,f_{\pm}(\Ncal\,)\, S_{l_\pm}\ ,
\end{equation}
where $f_{\pm}$ are two functions of the `total number operator'
\begin{equation}\label{tno}
\Ncal\ :=\ \frac{1}{2\th}\sum\limits^n_{a=1} z^a{\zb}^{\ab}
\qquad\textrm{satisfying}\qquad
f_{\pm}(r)\ =\ 0 \qquad\textrm{for}\quad r\le{l_\pm}{-}1 \ .
\end{equation}
The shift operators $S_{l_\pm}$ in (\ref{ans}) are defined to obey
\begin{equation}
S_{l_\pm}^\+\,S_{l_\pm}\ =\ 1 \qquad\textrm{while}\qquad 
S_{l_\pm}\,S_{l_\pm}^\+\=1-\Pi_{l_\pm}
\qquad\textrm{with}\qquad \Pi_{l_\pm} := 
\sum\limits_{|k|\le{l_\pm}-1}|k_1,\ldots,k_n\>\<k_1,\ldots,k_n| \ ,
\end{equation}
where $|k|:= k_1+\ldots +k_n$. Note that 
\begin{equation}
S_{l_\pm}^\+\,\Pi_{l_\pm}\=\Pi_{l_\pm}\,S_{l_\pm}\=0 \qquad\textrm{and}\qquad
f_{\pm}(\Ncal\,)\,\Pi_{l_\pm}\=\Pi_{l_\pm}\,f_{\pm}(\Ncal\,)\=0\ ,
\end{equation}
and $S_{l_\pm}^\+$ projects all states with $|k|<l_\pm$ out of $\cal H$.

One easily sees that (\ref{ans}) fulfills the homogeneous equations in
(\ref{mf1}) and (\ref{mf2}). Remembering that $\th_{a\bb}=-\th_{\bb
a}=\frac{1}{2\th^a}\,\de_{a\bb}
=\frac{1}{2\th}\,\de_{a\bb}$, we also obtain
\begin{align}\nonumber
\big[X^\pm_a\,,\, X^\pm_{\bb}\big]\ &=\ \th_{a\cb}\,\th_{\bb
  d}\,S_{l_\pm}^\+\,\bigl\{
f_{\pm}(\Ncal\,)\,{\zb}^{\cb}\,(1{-}\Pi_{l_\pm})\,z^d\,f_{\pm}
(\Ncal\,)\ -\ z^d\,f_{\pm}(\Ncal\,)\,(1{-}\Pi_{l_\pm})\,f_{\pm}(\Ncal\,)
\,\zb^{\cb}\bigr\}\,S_{l_\pm} \\[4pt]
\label{mf4}
&=\ -\sfrac{1}{4\th^2}\,\de_{a{\cb}}\,\de_{d{\bb}}\,S_{l_\pm}^\+\,\bigl\{
f_{\pm}^2(\Ncal\,)\,{\zb}^{\cb} z^d\ -\ f_{\pm}^2(\Ncal{-}1)\,z^d
{\zb}^{\cb}\bigr\}\,S_{l_\pm}
\end{align}
with the help of the identities
$\ {\zb}^{\cb}\,\Pi_{l_\pm}=\Pi_{l_\pm -1}\,{\zb}^{\cb}\ $ 
where $\ \Pi_0 := 0\ $ as well as
\begin{equation}
{{\zb}^{\cb}}\,f_{\pm}(\Ncal\,) \= f_{\pm}(\Ncal{+}1)\,{\zb^{\cb}}
\qquad\textrm{and}\qquad
{z^d}\,f_{\pm}(\Ncal\,) \= f_{\pm}(\Ncal{-}1)\,{z^d}\ .
\end{equation}

Substituting (\ref{mf4}) into (\ref{mf1}) and (\ref{mf2}), we employ
\begin{equation}
\de_{\cb d}\,z^d {\zb}^{\cb} \= 2\th\,\Ncal \qquad\textrm{and}\qquad
\de_{\cb d}\,{\zb}^{\cb} z^d \= 2\th\,(\Ncal+n)
\end{equation}
to find the conditions
\begin{align}\nonumber
0 &\= \de^{a\bb}\,\bigl[X^\pm_a\,,\,X^\pm_{\bb}\bigr]\ +\
\de^{a\bb}\,\th_{a\bb}\ \mp\ \frac{m}{4R^2} \\ \nonumber
&\=-\frac{1}{2\th}\,
S_{l_\pm}^\+\,\biggl\{ f_{\pm}^2(\Ncal\,)\,(\Ncal{+}n)\ -
\ f_{\pm}^2(\Ncal{-}1)\,
\Ncal\biggr\}\,S_{l_\pm}\ +\
\frac{n}{2\th}\ \mp\ \frac{m}{4R^2} \\ \label{mf5}
&\=\frac{1}{2\th}\,S_{l_\pm}^\+\,\biggl\{ \Ncal\,f_{\pm}^2(\Ncal{-}1)\ -\
(\Ncal{+}n)\,f_{\pm}^2(\Ncal\,)\ +\
n\,\Bigl(1 \mp \frac{m\,\th}{2nR^2}\Bigr)\biggr\}\,S_{l_\pm}
\end{align}
on the operators $f_{\pm}$. These recursions are solved by\footnote{
$\Theta$ denotes the Heaviside step function. 
It may be replaced by $1{-}\Pi_{l_{\pm}}$.
The ambiguity at $\Ncal=l_{\pm}{-}1$ is irrelevant here because then
the prefactor vanishes.}
\begin{equation}\label{fsol}
f_{\pm}^2(\Ncal\,)\= \Bigl(1 \mp \frac{m\,\th}{2nR^2}\Bigr)
\Bigl(1-\frac{{\cal Q}_{\pm}\ n!}{(\Ncal{+}1)\cdots(\Ncal{+}n)}\Bigr)\,
\Theta(\Ncal{-}l_{\pm}{+}1)\ ,
\end{equation}
where 
\begin{equation}\label{Qpm}
{\cal Q}_\pm\ :=\ \frac{l_\pm (l_\pm{+}1)\cdots(l_\pm{+}n{-}1)}{n!}\ .
\end{equation}
Assuming that $1-\frac{m\,\th}{2nR^2}>0$, we can take a positive
square root to get
\begin{equation}\label{Xsol}
X^{\pm}_a\=\frac{1}{2\th^\pm} \  S_{l_\pm}^\+\,\Bigl(1-
\frac{{\cal Q}_{\pm}\ n!}{(\Ncal{+}1)\cdots(\Ncal{+}n)}\Bigr)^{\frac{1}{2}}\,
\Theta(\Ncal{-}l_{\pm}{+}1)\,\de_{a\cb}\,{\zb}^{\cb}\, S_{l_\pm}\ ,
\end{equation}
where we abbreviated
\begin{equation}\label{thpm}
\th^\pm\ =\ \frac{\th}{\sqrt{1 \mp \frac{m\,\th}{2nR^2}}}\ .
\end{equation}
The $m{=}0$ case (equivalent to $\p{=}1$) is also covered by these solutions.
Note that for $m{\ne}0$ the solutions (\ref{Xsol}) coincide with those 
obtained in~\cite{Kraus,Nekrasov} if one assigns different 
noncommutative parameters $\th^+$ and $\th^-$ to the worldvolumes of 
D$(2n)$-branes and D$(2n)$-antibranes, respectively. 
Then the field strengths $F^\pm (\th^\pm )$ on ${\R}^{2n}_{\th^{\pm}}$
obtained from (\ref{Xsol}) will have finite topological charges ${\cal Q}_\pm$ 
given by (\ref{Qpm}), as calculated in~\cite{Kraus,Nekrasov}. 
The interesting idea of introducing distinct noncommutativity parameters on
multiple (coincident) D-branes (generated by different magnetic fluxes on
their worldvolumes~\cite{Dasgupta}) was discussed in~\cite{Tatar}.
This proposal gains support from our zero-tachyon BPS solutions (\ref{Xsol}) 
which carry oppositely oriented magnetic fluxes on branes versus antibranes.

\section{ Multi-instanton K-cycles }

In the remainder of this paper we will work towards clarifying the
D-brane interpretations of the multi-instanton solutions that we have
found. This will be done by illustrating that the solution
(\ref{ansatz1}), (\ref{ansatz2}) can be very naturally obtained via a
construction in K-homology. Passing from analytic to
topological K-homology will then provide a worldvolume picture of
these solutions in which the brane interpretations become manifest.
We shall find that, for $|m|=1$, a configuration of $k$
D$(2n{+}2)$-branes and $k$ D$(2n{+}2)$-antibranes wrapping a common sphere
$S^2$ with the monopole field is equivalent to $k$ D$(2n)$-branes and $k$
D$(2n)$-antibranes, i.e. the DUY equations on $\R^{2n}\times S^2$ are
equivalent to generalized vortex equations on $\R^{2n}$. This means
that instantons on $\rts$ are the spherical extensions of
vortices which are points in $\R^{2n}_\th$. Then, the solutions to the
generalized vortex equations produces (with $k{=}2^{n-1}$) $N_+$
D0-branes and $N_-$ D0-antibranes in the worldvolume $\R^{2n}_\th$. But
from the point of view of the initial brane-antibrane system on
$\rts$, they are spherical $N_+$ D2-branes and $N_-$
D2-antibranes. For $|m|>1$ this equivalence ceases to hold and requires
us to introduce the notion of a ``D-operation'' using another standard
construction in K-theory. In this case the solutions correspond
instead to D0-branes in the initial ${\rm D}(2n{+}2)-\overline{{\rm
D}(2n{+}2)}$ brane-antibrane system on $\rts$ which carry additional moduli
labelling their position in the auxiliary sphere $S^2$. In this
section we will describe the pertinent K-theoretic (co)cycles, and
then use them in the next section to illustrate these features.

\bigskip

\noindent
{\bf The monopole cocycle.} The existence of instanton solutions with
non-trivial flux relies crucially on the presence of a non-trivial
Dirac monopole configuration on the auxiliary space $S^2$, which is
also a crucial ingredient of the K-theoretic construction. The
K-theory charge group of the total space $\R^{2n}\times S^2$ can be
calculated through the suspension isomorphism to give
\begin{equation}
\K^0(\R^{2n}\times S^2)~=~\K^0(S^2)~=~\Z\oplus\Z \ ,
\label{K0susp}\end{equation}
where throughout K-theory with compact support is always
understood (i.e. $\R^{2n}$ is understood topologically via its
one-point compactification as $S^{2n}$). The Bott generator of the
reduced K-theory group $\widetilde{\K}^0(S^2)=\Z$ may be represented 
by the line bundle $\Lcal$ over $\C P^1\cong S^2$, which classifies the
Dirac monopole. Regarded as the non-trivial map between pairs of bundles 
over $\R^2$, i.e. as the class of a virtual bundle
$[\Lcal,\overline{\Lcal};\nu^{~}_{\rm D}]\in\widetilde{\K}^0(S^2)$, the
generator is determined by the standard (commutative) ABS configuration in
codimension~$2$,
\begin{equation}
(\nu^{~}_{\rm D})_{(r,\varphi)}\=\e^{\im\varphi} \ ,
\label{tauABS}\end{equation}
where $(r,\varphi)$ are polar coordinates on $\R^2$. This is the
canonical generator of $\pi_1(S^1)=\Z$ and is the usual commutative
tachyon field configuration in codimension~$2$.

Bott periodicity then induces an isomorphism
\begin{equation}
\alpha\,:\,\K^0(\R^{2n})\otimes\K^0(S^2)~\stackrel{\cup}
{\longrightarrow}~\K^0(\R^{2n}\times S^2)~\longrightarrow~
\K^0(\R^{2n}) \ ,
\label{Bottiso}\end{equation}
where the first map is the cup product and we use the fact that all
K-theory groups in the present case are freely generated. Explicitly,
evaluated on a virtual pair $[E^+,E^-;\tau]\in\K^0(\R^{2n})$, with
$\tau$ the tachyon field isomorphism on the rank $k$ bundles $E^-\to
E^+$ at infinity, the isomorphism (\ref{Bottiso}) is given by
\begin{equation}
\big[E^+\,,\,E^-\,;\,\tau\big]~\longmapsto~\alpha
\big[E^+\otimes(\Lcal\oplus\overline{\Lcal}\,)\,,\,
E^-\otimes(\Lcal\oplus\overline{\Lcal}\,)\,;\,\tau\bullet\nu^{~}_{\rm D}\big]
\label{Bottisoexpl}\end{equation}
with
\begin{equation}
\tau\bullet\nu^{~}_{\rm D}\=\begin{pmatrix}\tau\otimes\Idd_{2}&
\Idd_{k}\otimes\nu_{\rm D}^\dag\\[4pt]
\Idd_{k}\otimes\nu^{~}_{\rm D}&-\tau^\dag\otimes\Idd_{2}
\end{pmatrix} \ .
\label{Totimestau}\end{equation}
The peculiar form of this product owes to the periodicity properties
of the underlying Clifford algebra in the ABS construction, i.e. in
the standard decomposition of the gamma-matrices (\ref{G}) in terms of
lower-dimensional ones. For further details, see~\cite{OS1}. This
product isomorphism generated by the monopole bundle over $S^2$ will
play an important role in what follows. The topological equivalence
$\K^0(\R^{2n}\times S^2)=\K^0(\R^{2n})$ will essentially imply the
equivalence of the brane-antibrane systems on $\R^{2n}\times S^2$ and
$\R^{2n}$.

\bigskip

\noindent
{\bf Analytic tachyon K-cycles.} The configurations described by
(\ref{ansatz2}) correspond to noncommutative solitons. Written in
terms of the noncommutative ABS configuration (\ref{TT}), they
correspond in fact to noncommutative tachyons~\cite{Strom,HM1}. To
make this statement precise we will now show that these configurations are
directly related to ordinary commutative ABS configurations and can,
for our purposes, be simply treated by using the classical
constructions. This will also explicitly demonstrate that the K-theory
of the commutative and noncommutative configurations are the same,
which paves the way to our eventual worldvolume description. This is
accomplished via a standard K-theoretic mapping between analytic
(noncommutative) and topological (commutative)
descriptions~\cite{HM1,BD1}. The basic idea is that, viewed in the
noncommutative space $\rts$, the vortex configuration $\phi$
is an element of the algebra $C(S^2)\otimes{\cal K}(\C^k\otimes{\cal
  H})$, where $C(S^2)$ is the algebra of continuous complex-valued
functions on $S^2$ and $\Kcal$ denotes the algebra of compact
operators. By Morita equivalence, the corresponding
analytic K-theory classes live in
\begin{equation}
\K_0\big(C(S^2)\otimes{\cal K}(\C^k\otimes{\cal H})\big)~=~
\K_0\big(C(S^2)\big)~=~\K^0\big(S^2\big) \ ,
\label{analKoequiv}\end{equation}
which is the noncommutative version of the suspension isomorphism
(\ref{K0susp}). In this analytic setting, Bott periodicity
is the equivalence
\begin{equation}
\K_0\big(C(\rc)\otimes\cliff_2\big)\=\K_0\big(C(\rc)\big) \ .
\label{Bottanalytic}\end{equation}

The crux of this identification is the fact that the noncommutative
ABS configuration $T=T_1$ in (\ref{TT}) defines a cycle of the
analytic K-homology group $\Ka_0(\R^{2n})$~\cite{HM1,Sz1}. With
$\Hcal^\pm=\Delta^\pm\otimes\Hcal$, it is a bounded
Fredholm operator $T:\Hcal^-\to\Hcal^+$. Let $\Bcal(\Hcal^\pm)$ denote
the algebras of bounded linear operators on the separable Hilbert
spaces $\Hcal^\pm$, and let $\rho^\pm:C(\R^{2n})\to\Bcal(\Hcal^\pm)$
be representations of the algebra of functions on $\rt$ by pointwise, diagonal
multiplication (representing $\Hcal$ in the Bargmann polarization, for
example). Then for each $f\in C(\R^{2n})$, the operator $T$ is
``almost'' compatible with the two representations in the sense that
\begin{equation}
T\,\rho^-(f)-\rho^+(f)\,T~\in~\Kcal(\Hcal^+\oplus\Hcal^-) \ .
\label{Talmost}\end{equation}
In addition, we have
\begin{equation}
T\,T^\dag-\Idd~\in~\Kcal(\Hcal^+) \qquad\textrm{and}\qquad
T^\dag\,T-\Idd~\in~\Kcal(\Hcal^-) \ .
\label{TKcal}\end{equation}
Such a quintuple $(\Hcal^+,\Hcal^-,\rho^+,\rho^-;T)$ is called an
(even) Fredholm module. The abelian group of stable homotopy classes
of Fredholm modules is the analytic K-homology group
$\Ka_0(\R^{2n})=\K^0(C(\R^{2n}))$. The charge of a class
$[\Hcal^+,\Hcal^-,\rho^+,\rho^-;T]\in\Ka_0(\R^{2n})$ is the analytic
index of $T$,
\begin{equation}
{\rm index}(T)\=\dim\ker(T)\ -\ \dim\,{\rm coker}(T) \ .
\label{indexT}\end{equation}

The connection with the commutative K-theory description now proceeds
with the observation that the same K-homology class comes from a Dirac
operator on the space $\R^{2n}$~\cite{BD1}. For this, let us consider
the chiral spinor bundles $\mDelta^\pm\to\R^{2n}$ of rank $k=2^{n-1}$, and let
$\Dirac=-\im\gamma\cdot\partial:C^\infty(\rc,\mDelta^-)\to
C^\infty(\rc,\mDelta^+)$ be the corresponding Dirac operator on the
spaces of smooth spinors on $\rc$. By completing these
vector spaces using the induced inner product from the metric of
$\R^{2n}$, we may view $\Dirac$ as an unbounded linear operator
\begin{equation}
\Dirac\,:\,L^2(\rc,\mDelta^-)~\longrightarrow~L^2(\rc,\mDelta^+) \ .
\label{DiracL2}\end{equation}
With $\rho^\pm:C(\rc)\to\Bcal(L^2(\rc,\mDelta^\pm))$ the representations by
pointwise multiplication as above, it follows that
$(L^2(\rc,\mDelta^+),L^2(\rc,\mDelta^-),\rho^+,\rho^-;\Dirac/|\Dirac|)$
is a Fredholm module. The corresponding class $[\Dirac]\in\Ka_0(\rc)$
depends only on the original Dirac operator
$\Dirac:C^\infty(\rc,\mDelta^-)\to C^\infty(\rc,\mDelta^+)$.

The particularly noteworthy aspect of this correspondence~\cite{BD1}
is that the index (\ref{indexT}) of the noncommutative ABS
configuration coincides with that of the corresponding Dirac operator,
\begin{equation}
{\rm index}(T)~=~{\rm index}(\Dirac) \ .
\label{indexTDirac}\end{equation}
On the other hand, the index of
$\Dirac:C^\infty(\rc,\mDelta^-)\to C^\infty(\rc,\mDelta^+)$ is just
the virtual dimension (zeroth Chern number) of the index class
$[\ker(\Dirac),{\rm coker}(\Dirac);\Dirac]\in\K^0(\rc)$, so that
\begin{equation}
{\rm index}(T)~=~\ch_0\big(\ker(\Dirac)\ominus{\rm coker}(\Dirac)
\big) \ .
\label{indexvirtual}\end{equation}
This coincides with the K-theory charge of the Bott class
$[\mDelta^+,\mDelta^-;\mu]\in\K^0(\rc)$ given by the ABS construction,
where $\mu_x:\Delta^-\to\Delta^+$ is Clifford multiplication by
$x\in\rc$,
\begin{equation}
\mu_x\=\frac{\gamma\cdot x}{|x|} \ ,
\label{Cliffmult}\end{equation}
and it is the generator of $\pi_{2n-1}({\rm U}(k))=\Z$. In this way, the
analytic index of the noncommutative tachyon operator (\ref{indexT})
coincides with the winding number of the classical ABS tachyon field
(\ref{Cliffmult}) at infinity in $\rc$.

It can be shown that all classes in $\Ka_0(\rc)$ arise in this
way~\cite{BD1}. In particular, the above construction also identifies
the Poincar\'e duality isomorphism $\Ka_0(\rc)\cong\K^0(\rc)$ between
K-homology and K-theory through the map
$[\Hcal^+,\Hcal^-,\rho^+,\rho^-;T]\mapsto[\mDelta^+,\mDelta^-;\mu]$. It is in
this way that we will be able to use a commutative description in what
follows, with the ordinary ABS field (\ref{Cliffmult}). Everything
that we have said also easily generalizes to the higher degree
configurations $T^{~}_N=(T)^N$ used in (\ref{ansatz1}) and
(\ref{ansatz2}). By considering the Dirac operator corresponding to
the twisted spinor bundles
$\mDelta^\pm\otimes((\Lcal)^{m}\oplus(\,\overline{\Lcal}\,)^{m})$, we
can further twist the Fredholm operator (\ref{ansatz2}) by the class
$[(\Lcal)^{m},(\,\overline{\Lcal}\,)^{m};(\nu^{~}_{\rm D})^m]$ of the
degree $m$ monopole bundle over $S^2$. Given the monopole projectors
$P_\pm$ in (\ref{pro}), we choose a partial isometry $\varsigma$ on $\C^2$ obeying
\begin{equation}
\varsigma^\dag\,\varsigma\=\Idd_2\qquad\textrm{while}\qquad\varsigma\,
\varsigma^\dag\=\Idd_2-P_+ \ .
\label{varsigmaPdef}\end{equation}
Then from the relations $P^{~}_{N_\pm}=1-T^{~}_{N_\pm}T_{N_\pm}^\dag$
we may compute the corresponding index as
\begin{eqnarray}
{\rm index}\big(\phi\bullet(\varsigma)^m\big)
&=&\dim\ker\begin{pmatrix}\phi\otimes\Idd_2&\Idd_k\otimes
(\varsigma^\dag)^m\\[4pt]\Idd_k\otimes(\varsigma)^m&-\phi^\dag\otimes
\Idd_2\end{pmatrix}\ -\ \dim\ker\begin{pmatrix}\phi^\dag\otimes
\Idd_2&\Idd_k\otimes(\varsigma^\dag)^m\\[4pt]\Idd_k
\otimes(\varsigma)^m&-\phi\otimes\Idd_2\end{pmatrix}\nonumber\\[4pt]&=&
m\,\dim\ker(T_{N_-}^\dag)\ -\ m\,\dim\ker(T_{N_+}^\dag)
\nonumber\\[4pt]&=&m\,\Tr^{~}_{\C^k\otimes\Hcal}(P^{~}_{N_-})\
-\ m\,\Tr^{~}_{\C^k\otimes\Hcal}(P^{~}_{N_+})~=~m\,(N_--N_+) \ .
\label{indexphi}\end{eqnarray}
The K-theory charge (\ref{indexphi}) of the noncommutative soliton
configuration (\ref{ansatz1}), (\ref{ansatz2}) thereby coincides with
the topological charge $-{\cal Q}$ computed in the Yang-Mills theory on
$\rts$ in (\ref{TopchargeYM}).

\section{Worldvolume interpretation}

Using the analysis of the previous section we are now ready for the D-brane
interpretation of the multi-instanton solutions. For our purposes a
D-brane will be specified by a triple $(W,E,\sigma)$, where
$\R^1\times W\cong\R^1\times\R^{2n}$ is the brane worldvolume regarded
as a spin$^c$ submanifold of the ambient space $\R^1\times
X\cong\R^1\times W\times S^2$, and $\sigma:W\hookrightarrow X$ is its
embedding.\footnote{ In this context we are working with Type~IIA
  branes. More generally, we can take $W=\R^{2n}\times W_r$ where
  $W_r$ is any spin$^c$ manifold of dimension $r$ (with all fields
  independent of the coordinates on $W_r$). For odd $r$ this then
  allows for Type~IIB branes and degree~$1$ K-groups. For notational
  simplicity we only work explicitly with the case $W=\R^{2n}$, as it
  captures the essential features and is easily generalized.} $E\to
W$ is the complex Chan-Paton vector bundle with connection (extended
trivially along the time direction $\R^1$). More precisely, in the
following construction we will only need the stable isomorphism class
$\xi\in\K^0(W)$ of the Chan-Paton bundle. Virtual bundles will also be
encountered later on and correspond to (unstable) brane-antibrane
configurations wrapping $W$.

With a suitable equivalence relation put on $(W,E,\sigma)$, to be
described in detail below, the set of all equivalence classes of
D-branes $[W,E,\sigma]$ generates an abelian group called the
topological K-homology group $\Kt_0(X)$~\cite{HM1,Sz1}. The connection with the
noncommutative tachyon configurations of the previous section is
provided by the isomorphism
\begin{equation}
\kappa\,:\,\Kt_0(X)~\longrightarrow~\Ka_0(X)
\label{kappaiso}\end{equation}
of abelian groups which is defined as follows. Using the given data,
we form the Hilbert spaces $\Hcal_E^\pm=L^2(W,\mDelta^\pm(W)\otimes E)$, where
$\mDelta^\pm(W)\to W$ are the chiral spinor bundles of rank
$k=2^{n-1}$ induced by the spin$^c$ structure on $W$. Let
$\Dirac_E:\Hcal_E^-\to\Hcal_E^+$ be the corresponding twisted Dirac
operator. Then, as explained before, this construction produces an
analytic K-cycle which defines a class $[\Dirac_E]\in\Ka_0(W)$. Under
the push-forward $\sigma_*:\Ka_0(W)\to\Ka_0(X)$ this induces an
analytic K-homology class on the ambient space $X$ and the map
(\ref{kappaiso}) is thereby defined by
\begin{equation}
\kappa[W,E,\sigma]\=\sigma_*[\Dirac_E] \ .
\label{kappadef}\end{equation}
It can be shown that this map is well-defined and invertible~\cite{BD1}. This
follows essentially from our previous remarks about the equivalence
between the commutative and noncommutative descriptions.

There are three equivalence relations that need to be put on the
D-brane $(W,E,\sigma)$~\cite{Sz1}. They are bordism (continuous deformations of
the brane worldvolume $W$ and of the Chan-Paton bundle $E$), direct
sum (gauge symmetry enhancement for coincident D-branes), and vector
bundle modification (dielectric effect). We claim that for $|m|=1$ the
multi-instanton solutions (\ref{ansatz1}), (\ref{ansatz2}) provide a
physical realization of this latter relation, so we shall study it in
some detail.

Starting from $(W,E,\sigma)$ we can build another triple which
represents the same K-homology class by using the Baum-Douglas clutching
construction~\cite{BD1}. For this, let $I=W\times\C$ denote the trivial line
bundle over $W$, and let $H\to W$ be a spin$^c$ vector
bundle of rank $2$. We use the induced metric on $H$ to
define the ball and sphere bundles $\mball(H)$ and $\msphere(H)$
over $H$, which are also spin$^c$ vector bundles. In particular, the
sphere bundle of $H\oplus I$ can be constructed by gluing together
two copies $\mball(H)_\pm$ of the ball bundle of $H$ using the
identity map along their common boundary
$\msphere(H)=\partial\mball(H)_\pm$ to give
\begin{equation}
\msphere(H\oplus I)~=~\mball(H)_+~\cup^{~}_{\msphere(H)}~\mball(H)_- \ .
\label{msphereHI1}\end{equation}
Note that locally the sphere bundle $\msphere(H\oplus I)$ is
isomorphic to our ambient space $X=W\times S^2$ and so, with a slight
abuse of notation, we will identify these two manifolds. Let
\begin{equation}
\pi\,:\,\msphere(H\oplus I)~\longrightarrow~W
\label{piproj}\end{equation}
be the corresponding bundle projection.

Denote by $\mDelta^\pm(H)$ the pull-backs of
the chiral spinor bundles $\mDelta^\pm(W)\to W$ to $H$. As usual, the
vector bundle map induced by Clifford multiplication
\begin{equation}
\mu\,:\,\mDelta^-(H)~\longrightarrow~\mDelta^+(H)
\label{CliffDeltaH}\end{equation}
is an isomorphism off the zero section of $H\to W$, i.e. at
``infinity''. We may now define a vector bundle over (\ref{msphereHI1})
by putting $\mDelta^\pm(H)$ over $\mball(H)_\pm$. Since the line
bundles $\mDelta^\pm(H)\to\mball(H)$ are isomorphic
over $\msphere(H)$ by Clifford multiplication (\ref{CliffDeltaH}), we
can glue them together along the sphere bundle $\msphere(H)$ using the
transition function $\mu$ to define
\begin{equation}
\Xi~=~\mDelta^+(H)~\cup^{~}_{\mu|_{\msphere(H)}}~\mDelta^-(H) \ .
\label{Xidef}\end{equation}
This is the virtual bundle $\Xi=\mDelta^+(H)\ominus\mDelta^-(H)$ which
defines a class $[\mDelta^+(H),\mDelta^-(H);\mu]\in\K^0(X)$. For each
point $w\in W$ on the D-brane worldvolume, $\pi^{-1}(w)$ is a
two-dimensional sphere $S^{2}$, and $\Xi|_{\pi^{-1}(w)}$ yields the
Bott class $[\mDelta^+,\mDelta^-;\mu]$ which generates
$\widetilde{\K}^0(S^2)=\Z$. This class is the same as that of the
monopole cocycle $[\Lcal,\overline{\Lcal};\nu^{~}_{\rm D}]$ introduced
previously.

Vector bundle modification is then the equivalence relation which can
be formulated as the equality between topological K-homology classes of
the D-branes
\begin{equation}
[W\,,\,\xi\,,\,\sigma]~=~[X\,,\,\Xi\otimes\pi^*\xi\,,\,\sigma\circ\pi]
\label{vecmoddef}\end{equation}
for any (virtual) Chan-Paton bundle $\xi\in\K^0(W)$. In particular, if
the left-hand side of (\ref{vecmoddef}) corresponds to the class of
$k$ ${\rm D}(2n)-\overline{{\rm D}(2n)}$ pairs wrapping $W\cong\rc$,
i.e. $\xi=E^+\ominus E^-$ with $\ch_0(E^\pm)=k$, then the right-hand
side corresponds to $k$ ${\rm D}(2n{+}2)-\overline{{\rm D}(2n{+}2)}$ pairs
wrapping $X\cong\rns$. This is simply the equivalence between
instantons on $\rns$ and vortices on $\rc$ that we encountered before.

The key point now is that for a suitable choice of tachyon field on
the left-hand side of (\ref{vecmoddef}), the right-hand side
coincides with the K-homology class of the
multi-instanton solution (\ref{ansatz1}), (\ref{ansatz2}). For this,
we recall that the classical ABS configuration (\ref{Cliffmult}) for
$x\in\rc$ represents the class of the noncommutative ABS operator $T$ in
(\ref{TT}). It follows that the noncommutative tachyon field
(\ref{ansatz2}) corresponds to the K-theory class
\begin{equation}
\xi\=\big[E^+\,,\,E^-\,;\,(\mu)^{N_+}\,(\mu^\dag)^{N_-}\big]
\label{ximultiexpl}\end{equation}
over the worldvolume $W\cong\rc$. On the other hand, the relation
(\ref{vecmoddef}) equates the resulting K-homology class with that
defined by
\begin{equation}
\Xi\otimes\pi^*\xi\=\big[\pi^*E^+\otimes(\Lcal\oplus\overline{\Lcal}\,)
\,,\,\pi^*E^-\otimes(\Lcal\oplus\overline{\Lcal}\,)\,;\,\widetilde{\mu}
\,\big]
\label{Ximpixi}\end{equation}
over the ambient space $X\cong\rns$, where
\begin{equation}
\widetilde{\mu}\=\begin{pmatrix}\pi^*\,(\mu)^{N_+}\,(\mu^\dag)^{N_-}\,\sigma^*
  \otimes\Idd_2&\Idd_k\otimes\nu_{\rm D}^\dag\\[4pt]
\Idd_k\otimes\nu^{~}_{\rm D}&-\pi^*\,(\mu)^{N_-}\,
(\mu^\dag)^{N_+}\,\sigma^*\otimes\Idd_2\end{pmatrix} \ .
\label{Ximpixiexpl}\end{equation}

The charge of the class (\ref{Ximpixi}) is given by (\ref{indexphi}) for 
$|m|=1$, and it thereby describes, through the standard
process of tachyon condensation on the unstable system of $k$
D$(2n{+}2)$-branes and $k$ D$(2n{+}2)$-antibranes wrapping $X$, a
configuration of spherical $N_+$ D$2$-branes and $N_-$
D$2$-antibranes. On the left-hand side of (\ref{vecmoddef}), these are
instead D0-branes arising from vortices left over from condensation in
the transverse space $\rc$. Thus the vortices become instantons with
worldvolume $S^2$. The worldvolume interpretation of the multi-instanton
solutions
corresponding to monopole charges $|m|>1$ is not so
straightforward. In this case one needs to twist
$\pi^*\xi$ instead with the higher-degree monopole bundle
$(\Lcal)^{m}$, as in (\ref{indexphi}). This twisting no longer
preserves the topological K-homology classes. We shall see below how to
overcome this difficulty and hence interpret these solutions
K-theoretically.

\bigskip

\noindent
{\bf D-brane charge.} We will now derive a geometrical formula for the
topological charge of the multi-soliton configuration in terms of the standard
characteristic classes associated with the configuration of
D-branes. For this, we regard the original worldvolume embedding as
the closed embedding $\sigma:W\to\msphere(H\oplus I)$, which has
normal bundle $\mball(H)_+-\msphere(H)\cong H$. The corresponding Gysin
homomorphism $\sigma_!:\K^0(W)\to\K^0(X)$ is then defined such that
the following diagram commutes:
\begin{equation}\begin{array}{ccc}
\K^0(W)&\longrightarrow&\Kt_0(W)\\& &\\{\scriptstyle\sigma_!}\,
\downarrow& &\downarrow\,{\scriptstyle\sigma_*}\\& &\\
\K^0(X)&\longleftarrow&\Kt_0(X)\end{array}
\label{Gysindef}\end{equation}
where the horizontal arrows denote Poincar\'e duality
isomorphisms. It may be conveniently represented in a way that
refers only to K-theory groups through the sequence of maps
\begin{equation}
\sigma_!\,:\,\K^0\big(W\big)~\longrightarrow~\K^0\big(H\big)~
\longrightarrow~\K^0\big(X\,,\,\mball(H)_-\big)~\longrightarrow~
\K^0\big(X\big) \ ,
\label{Gysinalt}\end{equation}
where the first arrow is the Thom isomorphism of $H$, the second one
is the excision isomorphism, and the last map is restriction induced
by the inclusion $(X,\emptyset)\hookrightarrow(X,\mball(H)_-)$. From
(\ref{Gysinalt}) one can then show~\cite{Jakob1} that
$\Xi\otimes\pi^*\xi=\sigma_!\xi\oplus\pi^*\xi$ in $\K^0(X)$, and the
summand $\pi^*\xi$ can be eliminated by using the gauge symmetry
enhancement relation when defining the topological K-homology
group. In other words, we can replace the vector bundle modification
relation (\ref{vecmoddef}) by
\begin{equation}
[W\,,\,\xi\,,\,\sigma]~=~[X\,,\,\sigma_!\xi\,,\,\sigma\circ\pi] \ .
\label{vecmodalt}\end{equation}

The Chern character of a K-cycle $(W,\xi,\sigma)$ in the homology of
the space $X$ is given by~\cite{BD1} the push-forward $\sigma_*$ of the
Poincar\'e dual of the characteristic class $\ch(\xi)\wedge{\rm Td}(TW)$
in the rational cohomology of the worldvolume $W$, where $\ch$ denotes
the (graded) Chern character and ${\rm Td}(TW)$ is the Todd class of
the tangent bundle of $W$. By using (\ref{indexTDirac}),
(\ref{kappadef}), (\ref{vecmodalt}) and the ordinary Atiyah-Singer
index theorem (expressed within the framework of K-homology~\cite{BD1}), we
thereby arrive at the topological formula
\begin{equation}
{\cal Q}_{|m|=1}~=~-{\rm index}(\phi)~=~-\int_W
\sigma^*\big(\ch(\sigma_!\xi)\wedge{\rm Td}(TX)\big) \ .
\label{Qindex}\end{equation}
This formula illustrates two important features. First of all, it
expresses the fact that the equivalence of the charges in the
commutative and noncommutative theories is simply the equality of the
analytic and topological indices. Secondly, it expresses the
topological charges of the multi-instantons on $\rts$ as the standard
formula~\cite{OS1} for the charge of a D-brane $(W,\xi,\sigma)$ in
terms of characteristic classes of the ambient space $X$. In
particular, it explicitly illustrates how the instanton number on
$\rns$ is equivalent to a vortex charge on $\rc$.

\bigskip

\noindent
{\bf D-operations.} The solutions we have obtained for $|m|>1$ possess
many properties different from those at $|m|=1$. For instance, in the
BPS case, instead of a simple holomorphicity or anti-holomorphicity
constraint as in (\ref{e1}) or (\ref{e2}) for $|m|=1$, the tachyon
field $\phi$ is required to be covariantly constant as in
(\ref{e1e2}). Furthermore, a major difference between the $|m|=1$ and
$|m|>1$ cases lies in their behaviour under the SU(2) isometry group
of the two-sphere. As we have discussed, for $|m|=1$ the gauge
potential has a generalized (i.e. up to gauge transformations)
SU(2) invariance. This invariance leads to the equivalence of
instantons on $\rns$ and vortices on $\rc$ since no additional moduli
arise from the $S^2$ dependence in this case. From the K-homology
point of view, this is equivalent to the vector bundle modification
argument given above. On the other hand, for $|m|>1$ the configuration
is no longer homogeneous over the $S^2$, so that rotations produce
moduli which are not spurious because they cannot be killed by gauge
transformations.

Therefore, our $|m|>1$ solutions should not be interpreted as
D2-branes in a D$(2n{+}2)$ brane-antibrane system, but rather as
D0-branes inside D$(2n{+}2)$ brane-antibrane pairs which may carry
additional moduli indicating their location not only in $\rc$ 
but also in $S^2$. For $|m|=1$ the $S^2$ moduli are unphysical because 
they are gauge artifacts, but this is not so for $|m|>1$. Thus the 
D$(2n)$ and D$(2n{+}2)$ brane systems are not equivalent, and the vector 
bundle modification argument given above breaks down. This moduli 
dependence is particularly clear from the form of the field strength 
tensors (\ref{cfabb}) and (\ref{cfvtvp}) which show how the monopole 
degrees of freedom interlace with the field theory degrees of freedom 
on $\rt$. For example, consider the zero tachyon sector where
(\ref{rduy1}) and (\ref{rduy2}) are satisfied. The $m$-monopole flux
on $S^2$ defines the degree $m$ line bundle $(\Lcal)^m\to S^2$. Then
eqs.~(\ref{rduy1}) and (\ref{rduy2}) imply that the degrees (first
Chern numbers) of the rank $k$ complex vector bundles $E^+$ and $E^-$
on the branes and antibranes are $m$ and $-m$, respectively (we
implicitly use here an appropriate compactification of $\rc$ to make
the total magnetic flux finite). Thus the abelian fluxes over $\rc$
and $S^2$ are correlated, and they lead to well-defined solutions with
finite action and topological charge. In the remainder of this section
we propose a K-theoretic interpretation which naturally explains this
flux correlation and which provides the appropriate extension of the
framework described above to the case $|m|>1$.

The mapping which mixes fluxes in the manner described above will be
referred to as a ``D-operation''.\footnote{This is not to be confused
  with the Steenrod square cohomology operations used in~\cite{DMW1}.}
An operation in K-theory is a natural map
\begin{equation}
\Psi\,:\,\K^0(W)~\longrightarrow~\K^0(W)
\label{Psigen}\end{equation}
defined for every worldvolume $W$, which is also natural in $W$. In
this sense an operation is a symmetry of K-theory. The only operations
in complex K-theory which are ring homomorphisms, i.e. which obey
$\Psi(\xi\oplus\xi'\,)=\Psi(\xi)\oplus\Psi(\xi'\,)$ and
$\Psi(\xi\otimes\xi'\,)=\Psi(\xi)\otimes\Psi(\xi'\,)$, are the Adams
operations. For each $m\in\Z$, they are given by
\begin{equation}
\Psi^m(E)~=~Q_m(\wedge^1E,\dots,\wedge^mE) \ ,
\label{Psimdef}\end{equation}
where $\wedge^pE$ denotes the (class of the) $p$-th exterior power of
the Chan-Paton bundle $E\to W$. Here $Q_0:=1$ and $Q_m$, $m\geq1$ is
the $m$-th Newton polynomial which expresses the symmetric function
$\sum_a(u_a)^m$ as the unique polynomial of the elementary symmetric
functions $e_p$ of $u_1,\dots,u_m$, i.e.
\begin{equation}
Q_m(e_1,\dots,e_m)\=\sum_{a=1}^m\,(u_a)^m\qquad\textrm{with}
\qquad e_p\=\sum_{a_1<\dots<a_p}u_{a_1}\cdots u_{a_p} \ .
\label{Newtondef}\end{equation}
For example,
\begin{equation}
Q_1(e_1)\=e_1\quad,\qquad Q_2(e_1,e_2)
\=(e_1)^2-2e_2\quad,\qquad Q_3(e_1,e_2,
e_3)\=(e_1)^3-3e_1e_2+3e_3\quad,
\label{Newton1stfew}\end{equation}
and so on. For $m<0$, (\ref{Psimdef}) is defined with the arguments
$\wedge^1\overline{E},\dots,\wedge^{|m|}\overline{E}$. The Adams
operations (\ref{Psimdef}) may be conveniently represented through the
generating function defined by
\begin{equation}
\sum_{m=1}^\infty\,(-t)^{m-1}~\Psi^m(E)\=
\frac\diff{\diff t}\ln\Bigl(1\ +\ \sum_{p=1}^\infty\,t^p\,\wedge^p
E\Bigr) \ .
\label{Adamsgenfn}\end{equation}

For our purposes the importance of the Adams operations stems from two
important properties that they possess (see~\cite{Karoubi1} for
further details). First of all, if $\cal L$ is (the class of) any line
bundle, then
\begin{equation}
\Psi^m({\cal L})\={\cal L}^m \ .
\label{PsimcalL}\end{equation}
Secondly, if $W=\rc$ and $\xi\in\K^0(W)$, then
\begin{equation}
\Psi^m(\xi)\=m^n\,\xi~:=~\underbrace{\xi\oplus\dots\oplus\xi}_{m^n} \ .
\label{Psimrc}\end{equation}
Let us now apply this symmetry to the pertinent bundle
$\xi\otimes\Lcal$ over $\rns$ used in
(\ref{ximultiexpl})--(\ref{Ximpixiexpl}). Using the fact that the
Adams operation is a ring homomorphism on K-theory, for a fixed
monopole charge $m\in\Z$ we find
\begin{equation}
\Psi^m(\xi\otimes\Lcal)\=\Psi^m(\xi)\otimes\Psi^m(\Lcal)\=(m^n\,\xi)
\otimes(\Lcal)^m \ .
\label{PsimxiLcal}\end{equation}
Now let $\Q_{\,m}$ be the subring of $\Q$ consisting of fractions with
denominator a power of $m$. We then define an operation
\begin{equation}
\widetilde{\Psi}^m\,:\,\K^0(W)~\longrightarrow~\K^0(W)\otimes\Q_{\,m}
\label{wtPsimdef}\end{equation}
by $\widetilde{\Psi}^m=\Psi^m/m^{n-1}$ for $W=\rc$. From
(\ref{PsimxiLcal}) it then follows that the corresponding Chern
character is modified to
\begin{equation}
\ch\big(\,\widetilde\Psi^m(\xi\otimes\Lcal)\big)\=m~\ch\big(\xi\otimes
(\Lcal)^m\big) \ .
\label{chPsim}\end{equation}

Consequently, the mapping $\xi\mapsto\widetilde{\Psi}^m(\xi)$ induces
the appropriate charge twisting as required in (\ref{TopchargeYM}) and
(\ref{indexphi}) by effectively increasing the number of D0-branes
left over after tachyon condensation by a factor of $|m|$. This
immediately leads to the physical implications of operations in
K-theory. We propose that the appropriate multi-instanton K-cycle
obtained from the reduction $\rns\to\rc$ is
$(W,\widetilde{\Psi}^m(\xi),\sigma)$, where $\ch_0(\xi)=N_--N_+$ as
before. From the point of view of the original noncommutative gauge
theory on $\rts$, it is the cycle
$(X,\widetilde{\Psi}^m(\Xi\otimes\pi^*\xi),\sigma\circ\pi)$, with the
appropriate higher-degree bundle (\ref{PsimxiLcal}) as specified by
our original ansatz (\ref{AA}) (note that the Adams operation obeys
the naturality condition $\Psi^m\,\pi^*=\pi^*\,\Psi^m$). Now, however,
since the Adams operation $\Psi^m$ is not generally an isomorphism on
K-theory (but rather only a symmetry),
the vector bundle modification relation (\ref{vecmoddef}) breaks down
and we no longer have the equivalence of the brane-antibrane systems
on $\rc$ and $\rns$. Note that for $|m|=1$, one has
$\widetilde{\Psi}^1(\xi)=\xi$, and the class of the D-brane remains as
before. In this way the (modified) Adams operation
$\xi\mapsto\widetilde{\Psi}^m(\xi)$ takes into account the
non-spurious monopole moduli dependence of the D-brane state for
$|m|>1$, and sharply captures the differences between the $|m|=1$ and
$|m|>1$ cases. More generally, we propose that the usual descent relations
among D-branes in string theory can be naturally understood as a
consequence of the symmetries of K-theory.

\vfill\eject

\section{Summary and discussion}

In this paper we have constructed explicit solutions to the Yang-Mills
equations on the noncommutative space $\rts$ with gauge group ${\rm
  U}(2k)$ and arbitrary magnetic flux over the $S^2$. We have obtained
generic BPS solutions corresponding to multi-instantons on $\rts$ by
solving the noncommutative Donaldson-Uhlenbeck-Yau equations. They are
uniquely determined by ${\rm U}(k)$ vortex configurations on
$\rt$. The BPS conditions on these solutions are given by (\ref{res})
or (\ref{res-}) and relate the noncommutativity parameters $\theta^a$,
the radius $R$ of the two-sphere, and the topological charge $m$ of
the monopole bundle in the space $\rts$. We have also obtained non-BPS
solutions to the full Yang-Mills equations on $\rts$ which likewise arise
from a reduction to $\rt$. Generally, the solutions are labelled by
three integers $(N_+,N_-,m)$ (with $N_+{=}0$ or $N_-{=}0$ in the BPS
case), and carry other moduli associated with the locations of the
noncommutative solitons in $\rc$. The moduli space of these solutions
is (up to discrete symmetries)
\begin{equation}
{\cal M}(N_+,N_-,m)\= \C^{nN_+}\times\C^{nN_-}\times\C P^1\ \subset\ 
{\rm Gr}^{~}_{N_+}(\infty)\times{\rm Gr}_{N_-}^{~}(\infty)\times\C P^1\ , 
\label{calMNm}\end{equation}
where ${\rm Gr}^{~}_N(\infty)$ is the infinite-dimensional
grassmannian manifold which parametrizes the rank-$N$ projectors on
the Hilbert space $\C^k\otimes\Hcal$, and the $\C P^1$ factor
parametrizes the position in $S^2$. This
manifold includes the special solutions which can be interpreted as
extrema of a string-inspired potential for the Higgs field
configurations on $\rt$. Note that neither the topological charge
(\ref{TopchargeYM}) nor the Yang-Mills energy (\ref{energyinst})
depend on the translational moduli in $\rc$.

Since ${\rm Gr}^{~}_N(\infty)$ is also the classifying space for rank
$N$ complex vector bundles, to each multi-instanton solution living in
(\ref{calMNm}) we may associate a (virtual) bundle
$\xi_0\otimes(\Lcal)^m$, where $\xi_0=E_0^+\ominus E_0^-$ and $E_0^\pm\to\rc$
are bundles of rank $N_\pm$. We interpret this as the Chan-Paton
bundle of $N_+$ D2-branes and $N_-$ D2-antibranes wrapping $S^2$,
inside a system of $k=2^{n-1}$ D$(2n{+}2)$ brane-antibrane pairs wrapping
$\rns$. This interpretation follows from the choice of ansatz for the
$u(2k)$-valued gauge potential $\ca$ on $\rts$. Generically, such a
gauge field would live on $2k$ coincident D-branes, all of which carry
the same orientation. However, with a choice of multi-monopole
solution on $S^2$, the gauge symmetry is broken down to $u(2k)\to
u(k)\oplus u(k)$. In the ansatz (\ref{AA}) we then couple the $u(k)$
gauge connection $A^+$ with the monopole projector and $A^-$ with the
anti-monopole projector, leading to opposite monopole charges between
pairs of D-branes and hence opposite orientations. In other words,
due to the presence of a non-trivial monopole field on the two-sphere,
one can alter the orientation of a D-brane such that different branes
from the initial set of $2k$ branes wrap with opposite orientation
around the $S^2$.

An interesting aspect of our solutions is the rather drastic
differences between the cases $|m|=1$ and $|m|>1$. This is even
apparent in the reduced form (\ref{energy}) of the Yang-Mills
functional on $\rt$. When $m^2=1$, the relative normalization between
the kinetic and potential energy terms for the Higgs field $\phi$ is
precisely of the form needed to rewrite the action as the Yang-Mills
functional of a superconnection, as is expected of the energy of a
basic brane-antibrane system (see e.g.~\cite{AIO1}). In this case BPS
solutions correspond to equating the action to the topological charge
$\cal Q$, which implies immediately the generalized vortex equations
that we found from the Donaldson-Uhlenbeck-Yau equations on $\rts$. On
the other hand, when $m^2>1$, the action (\ref{energy}) does not yield
the standard superconnection form of a brane-antibrane energy
functional. BPS solutions in this case require not only the generalized vortex
equations to be satisfied, but also the more rigid equation
$|D\phi|^2=0$. A heuristic way to understand the differences here from the
point of view of the effective Yang-Mills-Higgs system obtained after
reduction is to look at the case $n=1$, $k=1$ with $A^+=-A^-$ ($B=0$)
in the commutative limit $\theta=0$. Then the action (\ref{energy})
describes the standard Ginzburg-Landau model of
superconductivity~\cite{Jaffe}. The case $m^2=1$ corresponds to the
BPS case when there are no forces between the vortices. On the other
hand, for $m^2>1$, vortices attract each other and there is a bound
state with finite energy and topological charge (but it is not a BPS
solution as the second order Yang-Mills-Higgs equations are
solved). Using this analogy it is tempting to speculate that in these
instances some analog of the Meissner efffect (complete expulsion of
magnetic flux) occurs in the combined brane-monopole system on
$\rts$. From the K-theory point of view, the multi-instanton solutions
on $\rts$ correspond to a symmetry in K-theory acting on the initial
brane-antibrane Chan-Paton bundle. For $m^2=1$ this symmetry is simply
the identity map and it directly yields the equivalence of the
brane-antibrane systems on $\rt$ and $\rts$ in topological K-homology,
but this is not so in the cases $m^2>1$. It would be interesting to
understand in more detail the D-brane physics of the noncommutative
solitons constructed in this paper and to further clarify the role of
the Adams operations in their description.

\bigskip

\bigskip

\noindent
{\bf Acknowledgements}

This work was partially supported by the Deutsche Forschungsgemeinschaft 
(DFG). The work of R.J.S. was supported in part by an Advanced Fellowship 
from the Particle Physics and Astronomy Research Council~(U.K.).

\bigskip

\bigskip

\end{document}